\documentclass[]{aa}

\usepackage{graphicx}
\usepackage{dcolumn}
\usepackage{bm}
\usepackage{mathrsfs}
\usepackage{natbib}
\usepackage{amssymb}
\usepackage{amsmath}
\usepackage{textcomp}

\bibpunct{(}{)}{;}{a}{}{,}
\def\xspec{{\sc xspec}}

\def\ktbb{kT_{\rm bb}}
\def\kte{kT_{\rm e}}
\def\me{m_{\rm e}}
\def\ne{n_{\rm e}}
\def\bmc{{\sc bmc}}

\def\comptb{{\sc comptb}}
\def\modname{{\sc compmag}}
\def\gax {\ifmmode{_>\atop^{\sim}}\else{${_>\atop^{\sim}}$}\fi}  

\def\cm2{cm$^{-2}$}
\def\s1{s$^{-1}$}

\def\htau{h_{\tau}}

\def\r0{r_{\rm 0}}
\def\b0{\beta_{\rm 0}}
\def\Ec{E_{\rm c}}
\def\scw{Schwarzschild}
\def\b{\textbf}

\newcommand{\coeff}[3]{#1_{#2}^{#3}}

\newcommand{\tcoeff}[3]{\tilde #1_{#2}^{#3}}
\newcommand{\be}{\begin{equation}}
\newcommand{\ee}{\end{equation}}
\newcommand{\bea}{\begin{eqnarray}}
\newcommand{\eea}{\end{eqnarray}}

\begin{document}
\title{Numerical solution of the radiative transfer equation: X-ray spectral formation from cylindrical accretion 
onto a magnetized neutron star}
\author{R. Farinelli\inst{1,2}, C. Ceccobello\inst{2}, P. Romano\inst{1}, and L. Titarchuk\inst{2,3}}
\offprints{R. Farinelli, farinelli@ifc.inaf.it}

\institute{
INAF, Istituto di Astrofisica Spaziale e Fisica Cosmica, via U. La Malfa 153, 90146, Palermo, Italy
\and Dipartimento di Fisica, Universit\`a di Ferrara, via Saragat 1, 44100, Ferrara, Italy
\and NASA, Goddard Space Flight Center, Greenbelt, MD 20771, USA\\
}

\abstract{Predicting the emerging X-ray spectra in several astrophysical objects is of great importance, 
in particular when the observational data are compared with theoretical models. This 
requires developing numerical routines for the solution of the radiative transfer equation according to the expected physical conditions of the systems under study.}
{We have developed an algorithm solving the radiative transfer equation in the Fokker-Planck 
approximation when both thermal and bulk Comptonization take place. The algorithm
is essentially a relaxation method, where stable solutions are obtained when the system has reached its steady-state equilibrium.}
{We obtained the solution of the radiative transfer equation in the two-dimensional domain defined by the 
photon energy $E$ and optical depth of the system $\tau$ using finite-differences for the 
partial derivatives, and imposing specific boundary conditions for the solutions. We treated the case of cylindrical accretion onto a magnetized neutron star.
}
{We considered a blackbody seed spectrum of photons with  exponential distribution across the 
accretion column and for an accretion where the velocity reaches its maximum at the stellar surface and at the top of the accretion
column, respectively. In both cases higher values of the electron temperature and of the optical depth $\tau$
produce flatter and harder spectra. Other parameters contributing to the spectral formation are the steepness
of the vertical velocity profile, the albedo at the star surface, and the radius of the accretion column.
The latter parameter modifies the emerging spectra in a specular way for the two assumed accretion profiles.
} 
{The algorithm has been implemented in the \xspec\ package for \mbox{X-ray} spectral fitting and is specifically
dedicated to the physical framework of  accretion at the polar cap of a neutron star with a 
high magnetic field ($\ga 10^{12}$ G). This latter case is expected to be  typical of  
accreting systems such as \mbox{X-ray} pulsars and supergiant fast \mbox{X-ray} transients.}

\authorrunning{R. Farinelli et al. }
\titlerunning{Numerical solution of the radiative transfer equation}


\maketitle
\section{Introduction}
\label{intro}

The solution of the radiative transfer equation (RTE) that describes the modification of a seed photon spectrum due to Comptonization
in a plasma is a much debated mathematical problem.
The equation in its full form is indeed integro-differential \citep{pomraning73} and allows 
for analytical solutions  under some particular assumptions, such as electron temperature
$T_{\rm e}=0$ \citep{tz98} or in the energy domain when the emerging spectrum is a powerlaw \citep{tl95}.
If the photon energy exchange for scattering is low ($\Delta \nu/\nu \ll 1$), it is possible
to perform a Taylor expansion of the Comptonization operator around the photon initial energy, which
transforms the RTE from integro-differential to purely differential \citep{rl79}; this is known as the Fokker-Planck 
(FP) approximation. The necessary conditions to allow this mathematical approach are that the 
Compton-scattering process occurs below the Klein-Nishina regime, namely when the electron 
temperature $\kte$ is subrelativistic ($\la 100$ keV), and that 
the optical depth of the Comptonization region is $\tau \ga$ 1.
The regime of low temperature and high optical depth of the plasma indeed ensures that the spectrum is
almost isotropized, so that it is possible to use the Eddington approximation for the 
specific intensity $I(\nu)=J(\nu)+3\nabla\cdotp {\bf F(\nu)}$, where $J$ and ${\bf F}$ are the zero and first moment of the
intensity field, respectively.

Moreover, if the plasma is not static but subject to dynamical (bulk) motion with  velocity ${\bf v(\tau)}$, then
another condition for using the FP approximation is that $v(\tau)$ must be subrelativistic.
When all the above restrictions are considered, one obtains by computing the first two moments of 
the RTE  a main equation that describes the shape of the angle-averaged emerging intensity $J(\nu)$ 
of the Comptonization spectrum. 
The general form of the RTE for the photon occupation number $n(\nu)=J(\nu)/\nu^3$ with subrelativistic electron
temperature in the presence of bulk motion was first derived by \citet[][hereafter BP81]{bp81a}.
Later, \citet[][hereafter TMK97]{tmk97} showed that analytical solutions can be found if the
velocity profile (assuming spherical symmetry) of the matter follows the free-fall law, $v_R \propto R^{-1/2}$.
In this case, the equation can be written as $L_x n(x,\tau)+ L_{\tau} n(x,\tau)=-s(x,\tau)$ and the solution
can be obtained using the variable-separation method in the form
\begin{equation}
 n(x,\tau)=\sum_{k=1}^{\infty} c_k R_k(\tau) N_k(x),
\label{series}
\end{equation}

\noindent
where $c_k$ and $R_k(\tau)$ are the expansion coefficients and eigenfunctions of the space operator
$L_{\tau}$, respectively, while  $N_k(x)$ is the solution of the differential equation
\begin{equation}
 L_x N_k(x) - \gamma_k N_k(x)=-s(x),
\label{energy_op}
\end{equation}

\noindent
where $\gamma_k \propto \lambda^2_k$ and $\lambda^2_k$ is the k$^{\rm{th}}$-eigenvalue of the space operator.

The Comptonization spectrum is mostly dominated by the first eigenvalue (see Fig. 3 in TMK97),
while the terms $N_k(x)$ with $k \ge2$ represent the fraction of photons that leave the medium without
appreciable energy-exchange.
Starting from the results of TMK97, \citet[][hereafter F08]{f08} developed a model ({\sc comptb}) 
for the \mbox{X-ray} spectral fitting package \xspec, which computes the emerging spectrum by means of a numerical convolution of the Green's function of the energy operator with a blackbody (BB)-like seed spectrum.
The model has been successfully applied to a sample of low-mass \mbox{X-ray} binaries hosting a neutron star (NS).
The method of the variable separation has also been adopted by \citet[][hereafter BW07]{bw07}, to find analytical solutions of the
RTE in the case of cylindrical accretion onto the polar cap of a magnetized NS. The starting equation in BW07 is
formally the same as in BP81: the most significant difference is that the Thomson cross-section
is replaced by an angle-averaged cross-section that takes into account the presence of the
magnetic field ($B\sim 10^{12}$ G).
Following the results of \citet{ls82}, BW07 assumed a velocity profile $v(\tau) \propto - \tau$, which allowed
 the RTE to be separable in energy and space. Note that the assumed velocity profile implies that
the matter flow stagnates at the stellar surface, which is at odds with the solution of TMK97, where the matter velocity
is increasing towards the central object, which can be either a NS or a black hole (BH).
When the velocity profile  is not free fall-like (TMK97, F08), or  $\propto \tau$ (BW07), 
the variable separation method can no longer be applied and the solution of the RTE can be
obtained only with numerical methods.

We report a numerical algorithm that allows the solution of the RTE in the FP regime using
finite-differences for any desired velocity profile and seed photon spatial and energy distribution.
We apply in particular it to cylindrical accretion towards the polar caps of a magnetized NS, following the approach of BW07. The algorithm  essentially uses a  relaxation method, therefore it finds the asymptotic (stationary) solution of the RTE  for a given initial value (at time $t=0$) condition.
Our work is structured as follows: in Section \ref{algorithm} we describe the kernel of the 
algorithm for generic two-variable elliptic partial differential equations; in Section \ref{rte_general} we formulate the problem for the
general RTE and appropriate boundary conditions; in Section \ref{bw07_result} we consider
the more specific case of a system configuration with azimuthal symmetry, 
typical of cylindrical accretion; in Section \ref{results} 
we show the emerging spectra obtained for different sets of the theoretical parameter space; finally, in Section \ref{4xspec} we briefly discuss possible astrophysical consequences and implementations 
(e.g., for \xspec) derived from the application of the algorithm.

\section{The general elliptic partial differential equations\label{algorithm}}

The algorithm we report is essentially based on the relaxation methods, which allow one to find the solution of a boundary 
elliptical problem. 
The differential equation has to be written by finite differences. Once the sparse matrix is 
defined, it can be split into layers over which an iteration process is applied until a solution is found \citep{NR}.   
The general form of a linear second-order elliptic equation with 
vanishing mixed derivatives and a source term can be written as
\begin{eqnarray}
\mathcal P(x,y)\frac{\partial^2 u}{\partial x^2}+\mathcal Q(x,y)\frac{\partial u}{\partial x}+\mathcal R(x,y)u+\mathcal W(x,y)\frac{\partial^2 u}{\partial y^2}\nonumber\\
+\mathcal Z(x,y)\frac{\partial u}{\partial y}=\frac{\partial u}{\partial t}-\mathcal S(x,y). \label{elleqn}
\end{eqnarray}
We define a three-dimensional grid of discrete points for the variables $x,y$ and $t$;
\begin{eqnarray}
 & & x_i=x_0+ih_x\,,\,\,\,\,\,i=0,1,\dots,N_x, \nonumber\\
 & & y_j=y_0+jh_y\,,\,\,\,\,\,j=0,1,\dots,N_y, \nonumber\\
 & & t_m=t_0+mh_t\,,\,\,\,\,\,m=1,2,\dots,M,
\end{eqnarray}
where $h_x,h_y,h_t$ are the grid spacing. The function $u(x,y,t)$ is evaluated at any point of the grid, so we 
write it as $u_i^{j,m}$. We write the first and second derivatives over the variables using finite differences:
\begin{eqnarray}
 \frac{\partial u}{\partial x} &=&\frac{u_{i+1}^{j,m}-u_i^{j,m}}{h_x}\,,\,\,\,\,\,\frac{\partial^2 u}{\partial x^2}=\frac{u_{i+1}^{j,m}-2u_i^{j,m}+u_{i-1}^{j,m}}{h_x^2},\nonumber\\
 \frac{\partial u}{\partial y} &=&\frac{u_{i}^{j+1,m}-u_i^{j,m}}{h_y}\,,\,\,\,\,\,\frac{\partial^2 u}{\partial y^2}=\frac{u_{i}^{j+1,m}-2u_i^{j,m}+u_{i}^{j-1,m}}{h_y^2},\nonumber\\
 \frac{\partial u}{\partial t} &=&\frac{u_{i}^{j,m}-u_i^{j,m-1}}{h_t}.
\end{eqnarray}
Substituting the above definitions into equation (\ref{elleqn}) and collecting terms, we obtain  
\begin{eqnarray}
\label{geneqn}
 a_i^ju_{i-1}^{j,m}+b_i^ju_i^{j,m}+c_i^ju_{i+1}^{j,m}+d_i^ju_i^{j-1,m}+e_i^ju_i^{j,m}+f_i^ju_i^{j+1,m}\nonumber\\
=\frac{(u_i^{j,m}-u_i^{j,m-1})}{h_t}-S_i^j,\,\,\,\,\nonumber\\
\end{eqnarray}
where
\begin{eqnarray}
a_i^j &=& \frac{\mathcal P(x_i,y^j)}{h_x^2}, \nonumber\\
b_i^j &=& -\frac{2\mathcal P(x_i,y^j)}{h_x^2}-\frac{\mathcal Q(x_i,y^j)}{h_x}+\mathcal R(x_i,y^j),\nonumber\\
c_i^j &=& \frac{\mathcal P(x_i,y^j)}{h_x^2}+\frac{\mathcal Q(x_i,y^j)}{h_x}, \nonumber\\
d_i^j &=&\frac{\mathcal W(x_i,y^j)}{h_{y}^2}, \nonumber\\
e_i^j &=&-\frac{2\mathcal W(x_i,y^j)}{h_y^2}+\frac{\mathcal Z(x_i,y^j)}{h_y}, \nonumber\\
f_i^j &=&\frac{\mathcal W(x_i,y^j)}{h_y^2}-\frac{\mathcal Z(x_i,y^j)}{h_y},\nonumber\\
S_i^j &=& \mathcal S(x_i,y^j).\label{coeffs}
\end{eqnarray}
\noindent
The operators over the $x$ and $y$ variables are then defined as
\begin{eqnarray}
\Delta_{x} &=& a_i^j+b_i^j+c_i^j,\nonumber\\
\Delta_{y} &=& d_i^j+e_i^j+f_i^j.
\label{operators}
\end{eqnarray}
\noindent
The solution procedure consists of dividing equation (\ref{geneqn}) into two equations. The first gives us the solution for an intermediate $m-1/2$ layer, while the second provides the solution for the $m$-layer. 
As starting point we need to establish an initial guess for the function at $m=1$. The system of equations to be solved is thus
\begin{eqnarray}
 \Delta_{x}u^{m-1/2}+\Delta_{y}u^{m-1} &=& \frac{u^{m-1/2}-u^{m-1}}{h_t}-S,\nonumber\\
 \Delta_{y}(u^{m}-u^{m-1}) &=&\frac{u^{m}-u^{m-1/2}}{h_t}, \label{system}
\end{eqnarray}

\noindent
in which we have temporarily dropped the indices $i,j$.
From the system we notice that the intermediate layer $m-1/2$ is needed only to build up the solution at the subsequent layer in $m$. 
The numerical accuracy of the solution can be estimated by combining both equations in the system (\ref{system}), which gives
\begin{eqnarray}
 \Delta_{x}u^m+\Delta_{y} u^m=\frac{u^m-u^{m-1}}{h_t}-S\nonumber\\
+ h_t\Delta_{x}\Delta_{y}\left(u^m-u^{m-1}\right).\label{eq+err}
\end{eqnarray}
\noindent
The third term on the right-hand side of equation (\ref{eq+err}) represents the residual error in the numerical solution. 
As a first step, we collect the terms with the same index in both equations (\ref{system}) and obtain
\begin{eqnarray}
\label{final_system}
\left(\Delta_{x}-\frac{1}{h_t}\right)u^{m-1/2} &=& -\left(\Delta_{y}+\frac{1}{h_{t}}\right)u^{m-1}-S, \nonumber\\
\left(\Delta_{y}-\frac{1}{h_t}\right)u^{m} &=& \Delta_{y}u^{m-1}-\frac{u^{m-1/2}}{h_t}. 
\label{system_new}
\end{eqnarray}

\noindent
Both equations are defined inside a 2D $(x,y)$-domain, with boundary conditions defined according to the 
specific problem under consideration. 
First, for any $m$ and $j$ values, we must impose the boundary condition on the left-hand side of the $x$-domain
($i=0$) for the function $u_0^{j,m-1/2}$
\begin{eqnarray}
u_0^{j,m-1/2}=g^j_0,
\label{leftbc_energy}
\end{eqnarray}

\noindent
while the source term $S^j_0$ is defined at the beginning.
Thus, for $i=0$, equation (\ref{geneqn}) can be written as
\begin{eqnarray}
u_0^{j,m-1/2}= \hat{L}_0^ju_1^{j,m-1/2}+\hat{K}_0^{j},
\label{u_0_x}
\end{eqnarray}

\noindent
where
\begin{eqnarray}
\hat{L}_0^{j} &=&-\frac{c_0^j}{b_0^j-\frac{1}{h_t}}\,,\,\,\,\,\,\hat{K}_0^{j}=\frac{\hat S_0^{j,m-1}}{b_0^j-\frac{1}{h_t}}\nonumber\\
\hat{S}_0^{j,m-1} &=& -\left(d_0^j+e_0^j+f_0^j+\frac{1}{h_t}\right)u_0^{j,m-1}-S_0^j, \label{gen1eqn_coeffs0}
\end{eqnarray}

\noindent
with the coefficients determined in equation (\ref{coeffs}).
\noindent
For $i=1$, using equation (\ref{u_0_x}) we obtain
\begin{eqnarray}
 \left(a_1^j\hat L_0^{j}+b_1^j-\frac{1}{h_t}\right)u_1^{j,m-1/2}+c_1^ju_2^{j,m-1/2}\nonumber\\
=\hat S_1^{j,m-1}-a_1^j\hat K_0^{j},
\end{eqnarray}

\noindent
which can be written as 
\begin{equation}
 u_1^{j,m-1/2} = \hat L_1^{j}u_2^{j,m-1/2}+\hat K_1^{j},
\end{equation}

\noindent
where
\begin{equation}
\hat L_1^{j}=-\frac{c_1^j}{a_1^j\hat L_0^{j}+ b_1^j-\frac{1}{h_t}}\,,\,\,\,\,\hat K_1^{j}=\frac{\hat S_1^{j,m-1}-a_1^j \hat K_0^{j}}{a_1^j\hat L_0^{j}+b_1^j-\frac{1}{h_t}}.
\end{equation}

\noindent
Iterating the process, we obtain the general form
\begin{equation}
 u_i^{j,m-1/2}= \hat L_i^{j}u_{i+1}^{j,m-1/2}+\hat K_i^{j},
\label{gen1eqn}
\end{equation}

\noindent
where
\begin{eqnarray}
\hat L_i^{j} &=& -\frac{c_{i}^j}{a_i^j\hat L_{i-1}^{j}+ b_i^j-\frac{1}{h_t}}\,,\,\,\,\,\hat K_i^{j}=\frac{\hat S_i^{j,m-1}-a_i^j \hat K_{i-1}^{j}}{a_i^j\hat L_{i-1}^{j}+b_i^j-\frac{1}{h_t}}.\nonumber\\ \label{gen1eqn_coeffs}
\label{lhat_khat}
\end{eqnarray}
\noindent
At the right boundary of the $x$-domain ($i=N_x$), we impose the second boundary condition 
\begin{equation} 
\coeff{u}{N_x}{j,m-1/2}=\coeff{g}{N_x}{j}.
\label{rightbc_energy}
\end{equation} 
\noindent 
Now, using equation (\ref{gen1eqn}) we can thus build up the solution over the $x$-variable iteratively as  
\begin{eqnarray}
 u_{N_{x}-1}^{j,m-1/2} &= &\hat L_{N_{x}-1}^{j} g_{N_x}^{j} + \hat K_{N_x-1}^{j},\nonumber\\
 u_{N_{x}-2}^{j,m-1/2} &= &\hat L_{N_{x}-2}^{j} u_{N_x-1}^{j,m-1/2} + \hat K_{N_x-2}^{j},\nonumber\\
 \dots\,\,\,\,\,& \,&\,\,\,\,\dots\nonumber\\
 u_{0}^{j,m-1/2} &=&\hat L_{0}^{j} u_{1}^{j,m-1/2} + \hat K_{0}^{j}.
\end{eqnarray}
\noindent
Therefore, the construction of the solution is obtained in two steps:  a bottom-up process which
allows one to build the coefficients $\hat L_i^{j}$ and $\hat K_i^{j}$ (Eq. [\ref{lhat_khat}]) starting from
the left boundary condition on $u^j_0$ (Eq. [\ref{leftbc_energy}]), followed by a top-down procedure 
determined by the right boundary condition $\coeff{u}{N_x}{j}$ (Eq. [\ref{rightbc_energy}]).

Once the solution over the $x$-variable for the $m-1/2$ layer is obtained for any $j$ 
(the index of the $y$ variable), we then seek the solution of the second equation
in the system (\ref{system}) by following the same procedure described above with initial 
 boundary condition at $j=0$
\begin{equation}
u_i^{0,m}=\tilde{L}_i^0 u_i^{1,m}+\tilde{K}_i^{0},
\label{u_0_y}
\end{equation}

\noindent
and, similarly to equation (\ref{lhat_khat}) 
\begin{eqnarray}
\tilde L_i^{j} &=&-\frac{f_i^j}{d_i^j\tilde L_i^{j-i}+e_i^0+\frac{1}{h_t}}\,,\,\,\,\,\,\tilde K_i^{j} =\frac{\tilde S_i^{j,m}-d_i^j\tilde K_i^{j-1}}{d_i^j\tilde L_i^{j-i,m}+e_i^0+\frac{1}{h_t}},\nonumber\\
\end{eqnarray}

\noindent
where
\begin{equation}
\tilde S_i^{j,m} = \left(d_i^j+e_i^j+f_i^j+\frac{1}{h_t}\right)u_i^{j,m-1}-\frac{u_i^{j,m-1/2}}{h_t}, 
\end{equation}

\noindent
depends on the solutions $u_i^{j,m-1/2}$ and $u_i^{j,m-1}$ obtained in the layers $m-1/2$ and $m-1$.

\noindent
As required for the procedure over the $x$-variable, the coefficients $\tcoeff{L}{i}{j}$ and $\tcoeff{K}{i}{j}$,
built from  $\tcoeff{L}{i}{0}$ and $\tcoeff{K}{i}{0}$, are determined by the left boundary condition  ($j=0$) for the function $u^0_i$,
and the solution for any $j$ is determined by the right boundary condition  $\coeff{u}{i}{N_y}=\coeff{g}{i}{N_y}$:
\begin{eqnarray}
 \coeff{u}{i}{N_y-1, m} &=& \tcoeff{L}{i}{N_y-1}\coeff{u}{i}{N_y,m}+\tcoeff{K}{i}{N_y-1}, \nonumber\\
 \coeff{u}{i}{N_y-2, m} &=& \tcoeff{L}{i}{N_y-2}\coeff{u}{i}{N_y-1,m}+\tcoeff{K}{i}{N_y-2}, \nonumber\\
 \dots\,\,\,\,\,& \,&\,\,\,\,\dots\nonumber\\
 \coeff{u}{i}{0,m}\,\,\,\,\,\,\,\,\, &=&\tcoeff{L}{i}{0}\coeff{u}{i}{1,m}+\tcoeff{K}{i}{0}.
\end{eqnarray}
\noindent
After constructing the solution over the $x$ and $y$ variable, the solution of the top layer $m$ becomes
the initial function of the bottom layer related to iteration $m+1$ according to the scheme
\begin{eqnarray}
& m =&1 \rightarrow (u^0, u^{1/2}, u^1), \nonumber\\
& m =&2 \rightarrow (u^1, u^{3/2}, u^2), \nonumber\\
&\dots  &\dots\,\,\,\, \dots \nonumber\\
& m =&M \rightarrow (u^{M-1}, u^{M-1/2}, u^M).
\label{m_layers}
\end{eqnarray}
\noindent
It is also worth mentioning that at the first iteration $m=1$ an initial guess function $u^{j,0}_{i}$ must be
assumed, which is needed to find the solutions $u^{j,1/2}_{i}$ and $u^{j,1}_{i}$ in the
system (\ref{system}).
The loop over $m$ stops when the same convergence criterion is satisfied, which physically means that
the solution has "relaxed" to its stationary value. One possible criterion could be 
$1-\varepsilon < |u|^{m-1}|/|u|^{m-1/2} < 1+\varepsilon $ and $1-\varepsilon < |u|^{m-1/2}|/|u|^{m} < 1+\varepsilon $,
where $\varepsilon$ is a user-defined numerical tolerance. In the next section we will show instead another 
convergence criterion we have chosen for stopping the iteration procedure, for the particular case of the RTE.

\section{Application to the radiative transfer equation and boundary conditions\label{rte_general}}

The general form of the RTE in the presence of subrelativistic bulk motion for a plasma with constant temperature $T_{\rm e}$ is given 
by (see Eq.[18] in BP81)
\begin{eqnarray}
\frac{\partial n}{\partial t}+\b V\cdot\nabla n & =&\nabla\cdot\left(\frac{1}{3 \ne\sigma(\nu)}\nabla n\right)+\frac{1}{3}\left(\nabla\cdot \b V\right)\nu\frac{\partial n}{\partial \nu}\nonumber\\
& + & \frac{1}{\nu^2}\frac{\partial}{\partial \nu}\left[\frac{\ne \sigma(\nu)}{\me}\nu^4\left(n+T_e\frac{\partial n}{\partial \nu}\right)\right]+j(\nu, \b r),  \label{bpeqn}\nonumber\\
\nonumber\\
\label{bp81_eq}
\end{eqnarray}

\noindent
where $n(\nu, \bf{r})$ is the zero-moment occupation number of the radiation field intensity,  $\b V$ is the plasma bulk velocity vector,
$\sigma(\nu)$ is the electron scattering cross-section, $n_{\rm e}(\bf{r})$ is the electron density and $j(\nu, \b r)$ is the source term.
Because the spectral formation is determined by the optical depth $\tau$ of the system, we use the latter
quantity  as the actual space variable.
The solution of equation (\ref{bp81_eq}) is fulfilled by imposing the boundary condition at the surface defined by $\tau=0$, 
(which represents the starting point of the integration domain) for the spectral flux, which is given by
\begin{equation}
 \b F(\nu,r)= -\nu^3\left[\left(\frac{1}{3 \ne \sigma(\nu)}\nabla n\right)+\frac{1}{3}\b V\nu\frac{\partial n}{\partial \nu}\right].
\label{flux_nu}
\end{equation}
\noindent
Under particular symmetries of the system configuration (e.g., cylindrical or spherical), the problem
becomes one-dimensional.
For constant electron temperature $T_{\rm e}$ it is also more convenient to use the adimensional variable 
$x \equiv h\nu/\kte$; moreover, when performing numerical integration using finite-difference methods, 
we use a logarithmic binning of the energy through the additional change of variable $x \rightarrow e^q$.
Under these assumptions, equation (\ref{flux_nu}) becomes
\begin{equation}
F(q,\tau)= -\left[\frac{1}{3}\frac{\partial J}{\partial \tau}+\frac{1}{3} V \left(\frac{\partial J}{\partial q}-J\right)\right],
\label{flux_q}
\end{equation}

\noindent
where $J \equiv n~ x^3$ is the specific intensity.

\noindent
At the inner boundary we impose the condition

\begin{equation}
F(q,0)=-\frac{1}{2}\left(\frac{1-A}{1+A}\right)J,
\label{flux_bound}
\end{equation}

\noindent
where $A$ is the albedo at the surface. A fully absorptive surface ($A=0$) is appropriate for a BH, while $0 < A\leq1 $ accounts, e.g., for a NS atmosphere. 
However, the inner boundary condition (\ref{flux_bound}) depends on energy as well as  space (see Eqs. [\ref{flux_nu}] and [\ref{flux_q}]).
For mixed boundary value problems, no analytical solutions are possible
(see Appendix E in TMK97) and numerical methods prove to be unstable.
However, in the energy range where the spectrum is a powerlaw $J(x,\tau)=R(\tau) x^{-\alpha}$,
equation (\ref{flux_bound}) becomes
\begin{equation}
 -\frac{dR}{d\tau}+\beta_{\rm 0}(\alpha+3) R= -\frac{3}{2}\left(\frac{1-A}{1+A}\right) R,
\label{pl_bound}
\end{equation}

\noindent
where $\beta_{\rm 0}$ is the bulk velocity at the inner radius ($\tau=0$), and here the problem 
is reduced to a standard boundary condition over
the space variable $\tau$.
\noindent
Writing the derivative in terms of finite-difference, equation (\ref{pl_bound}) then becomes
\begin{equation}
-\frac{u^1_i-u^0_i}{\htau}+ \beta_{\rm 0}(\alpha+3) u^0_i= -\frac{3}{2}\left(\frac{1-A}{1+A}\right)u^0_i,
 \label{pl_bound_fd}
\end{equation}

\noindent
which can be written after collecting terms as
\begin{equation}
u^0_i=\frac{1}{1+\htau [\beta_{\rm 0} (\alpha+3)+ G(A)]}u^1_i, 
\label{leftbc_tau}
\end{equation}

\noindent
where $G(A)=3/2(1-A)/(1+A)$.
\noindent
We then set our problem as follows: first, because the solution $u^j_i$  of the RTE physically represents
a specific intensity, it must by definition be equal to zero in the limits $E\rightarrow 0$ and $E\rightarrow \infty$, therefore we set $u^j_0=u^j_{N_x}=0$ (see Eqs. [\ref{leftbc_energy}] and [\ref{rightbc_energy}]).
For the behavior of the function $u^j_i$ for $\tau=0$ ($j=0$), equation (\ref{leftbc_tau}) immediately
allows us to define (see Eq.~[\ref{u_0_y}])
\begin{equation}
\tcoeff{L}{i}{0}= \frac{1}{1+\htau [\beta_{\rm 0} (\alpha+3)+ G(A)]}, ~~~~~ \tcoeff{K}{i}{0}=0.
\label{ltilde_ktilde_j0}
\end{equation}
\noindent
As outer boundary condition over $\tau$, we impose that $u^{N_{y}}_i =0$, which means that the specific intensity
goes to zero for $\tau \rightarrow \tau_{\rm max}$.

\noindent
We emphasize that the condition $u^j_i > 0$ for any ($i, j$)-value implies a specific
restriction in the choice of the step size $h_{\tau}$, which ensures that  $\tcoeff{L}{i}{0} > 0$
(as $\beta_0 \le 0$). More specifically, we imposed the condition on $h_{\tau}$ such that
the number of steps over $\tau$ be $N_{\tau}=\tau_{\rm max}/h_{\tau} \ge 10$.

\subsection{The iteration procedure}

As already mentioned in Section 2,  it is necessary to choose a convergence criterion for stopping
the iteration over the $m$ variable.
We proceeded in the following way: at each iteration $m$, we computed the spectral index $\alpha_m$ of the 
solution $u^{0,m}_i$ (corresponding to $\tau=0$) in a given  energy range $E_{\rm min}-E_{\rm max}$.
To minimize  bias or wrong estimate of $\alpha_m$, the definition of the energy interval
for the computation of the spectral slope must be chosen carefully. 
If the seed photon spectrum is a BB with temperature $\ktbb$,  a reasonable choice can be
the assumption $E_{\rm min} \approx 7 \ktbb$ and $E_{\rm max} \approx 20 \ktbb$, respectively, given
that this interval is above the major contribution of the BB component and
below the expected high-energy  cut-off value.

Once  $\alpha_m$ is estimated, it is inserted into equation (\ref{ltilde_ktilde_j0}), which accordingly represents the boundary
condition at $\tau=0$ for the iteration $m+1$ (see Eq. [\ref{u_0_y}]). We then computed the new index $\alpha_{m+1}$ for
$u^{0,m+1}_i$, and again inserted it into equation (\ref{ltilde_ktilde_j0}) at iteration $m+2$, and so on.
The routine is stopped when $\alpha_m$ and $\alpha_{m+1}$ differ less than $10^{-5}$ provided that 
the condition holds for a sufficiently high number of iterations ($> 100$). Note that
the same criterion is adopted also if $\beta_0=0$, even if then 
of course $\tcoeff{L}{i}{0}$ remains constant across the iteration.
We have also verified that this criterion automatically also satisfies the convergence of
the norms  $|u|^{m-1}$, $|u|^{m-1/2}$, and $|u|^{m}$.

\section{Cylindrical accretion onto a magnetized neutron star\label{bw07_result}}

We applied our algorithm to solve the RTE for accretion towards the polar cap of a magnetized NS,
 whose mathematical formalism was developed by BW07 in the framework of the spectral formation of 
accretion-powered \mbox{X-ray} pulsars.
The relatively strong magnetic field ($B\ga 10^{12}\rm G$) of the NS is expected to channel the accretion flow
towards the polar caps, and  for low values of the altitude above the star surface, the
problem can be treated in a axis-symmetric approximation where the space variable is defined
by the vertical coordinate $Z$. 
The magnetic field moreover forces the medium to become birefringent as the effect of vacuum polarization,
and birefringence  entails the formation of two linearly polarized modes (ordinary and extraordinary) of 
the photons, each having a characteristic scattering cross-section.
For ordinary mode photons with energy below the first cyclotron harmonic at $E_{\rm c} \approx 11.57~B_{12}$ keV 
(where $B_{12}\equiv B/10^{12}$ G), BW07
defined angle-averaged cross-sections parallel and
perpendicular to the lines of the magnetic field as 
$\sigma_\parallel=10^{-3} \sigma_{\rm T}$ and $\sigma_\bot=\sigma_{\rm T}$,
respectively, where $\sigma_{\rm T}$  is the Thomson scattering cross-section.
This is indeed the only approximation that allows to treat the problem analytically or numerically.
We note that \cite{ferrigno09}, starting from the analytical solutions reported
in BW07, developed a model that was later almost successfully tested on the accreting pulsar \mbox{4U 0115+63}.
Their model is based essentially on the convolution of the column-integrated Green's function of the thermal plus bulk
scattering operator with a given seed photon distribution.
The basic assumption of this derivation is that the velocity profile of the accreting matter is assumed to be
$v(\tau) \propto -\tau$, which allows one to find analytical solutions through the variable separation
method (Eqs. [36] and [37] in BW07). The numerical algorithm we developed directly solves the RTE, 
without the need of this prescription for the dynamical configuration of the accreting matter
field, and we included some modifications with respect to the approach of BW07 and \cite{ferrigno09}.

First, following TMK97, we include in equation (\ref{bp81_eq}) a second term in the thermal Comptonization
operator that accounts for the contribution of the bulk motion velocity of electrons in addition to 
their thermal (Maxwellian) component.
\noindent
With this prescription in mind, equation (\ref{bp81_eq}) becomes

\begin{eqnarray}
\label{BW07eqn}
\hspace{-0.2truecm} \frac{1}{c}\frac{\partial n}{\partial t} &-& \mathcal S(\epsilon,Z) = 
-\frac{v}{c}\frac{\partial n}{\partial Z}
+\frac{dv}{dZ}\frac{\epsilon}{3c}\frac{\partial n}{\partial \epsilon}+\frac{\partial}{\partial Z}
\left(\frac{1}{3 \ne \sigma_\parallel}\frac{\partial n}{\partial Z}\right)\nonumber\\
- \frac{n}{t_{\rm{esc}}}&+ &\hspace{-0.2truecm}   \frac{\ne \overline\sigma}{\me c^2}\frac{1}
{\epsilon^2}\frac{\partial}{\partial \epsilon}\left[\epsilon^4\left(n+(\kte + \me v^2/3)
 \frac{\partial n}{\partial\epsilon}\right)\right], 
\end{eqnarray}

\noindent
where $\epsilon\equiv h\nu$, $\overline\sigma=10^{-1} \sigma_{\rm T}$, while
$t_{\rm{esc}}$ is the photon mean escape timescale (see Eq. [17] in BW07)
\begin{equation}
 t_{\rm{esc}}=\frac{\ne \sigma_\bot r^2_0}{c}.
\label{tesc}
\end{equation}

\noindent
Now, using the relation $d\tau=\ne \sigma_\parallel dZ$ and the logarithmic binning of the adimensional
energy $x \equiv h\nu/\kte$, equation (\ref{BW07eqn}) becomes

\begin{eqnarray}
\label{bweqn_aftersubs}
\frac{1}{\ne \sigma_\parallel c H}\frac{\partial J}{\partial t}-\frac{\mathcal S(q,\tau)}{H} = 
\left[1+\frac{\me v(\tau)^2}{3\kte}\right]\frac{\partial^2 J}{\partial q^2}\nonumber\\
+\left[\frac{3\kte(e^q-3+\hat{\delta})-{\me v(\tau)^2}}{3\kte}\right]\frac{\partial J}{\partial q}\nonumber\\
+ \left[e^q-3\hat{\delta}-\frac{\xi^2v(\tau)^2}{Hc^2}\right]J+\frac{1}{3H}
\frac{\partial ^2J}{\partial \tau^2}-\frac{v(\tau)}{Hc}\frac{\partial J}{\partial \tau},
\end{eqnarray}

\noindent
where we have defined the quantities
\begin{equation}
H= \frac{\overline \sigma}{\sigma_\parallel}\frac{\kte}{\me c^2},
\end{equation}

\noindent
and
\begin{equation}
 \hat{\delta}= \frac{1}{3H}\frac{d\beta(\tau)}{d\tau},
\end{equation}

\noindent
where $\beta(\tau)=v(\tau)/c$, while the dimensionless parameter $\xi$  is given by (see Eq. [26] in BW07)
\begin{equation}
\xi=\frac{15.8\, r_0}{\dot{m}}.
\label{csi_par}
\end{equation}

\noindent
Equation (\ref{bweqn_aftersubs}) is given in the general form (\ref{elleqn}) and for this particular case, we have

\begin{align}
\label{compactBW07}
&\mathcal P(\tau) = 1+\frac{\me v(\tau)^2}{3\kte},\nonumber\\
&\mathcal Q(\tau) = \frac{3\kte(e^q-3+\hat{\delta})-{\me v(\tau)^2}}{3\kte},\nonumber\\
&\mathcal R(\tau) = e^q-3\hat{\delta}-\frac{\xi^2 v(\tau)^2}{Hc^2},\nonumber\\
&\mathcal W(\tau) = \frac{1}{3H},\nonumber\\
&\mathcal Z(\tau) =-\frac{v(\tau)}{Hc},\nonumber\\
&\hat{\mathcal S}(q,\tau) = \frac{\mathcal S(q,\tau)}{H}. 
\end{align}

\noindent
To solve equation (\ref{bweqn_aftersubs}), it is necessary to define the behavior of the
velocity profile $\beta(\tau)$.
We considered two possibilities: in the first one, we assumed a general form 
\begin{equation}
\beta(Z)=-\mathscr{A}(Z_s/Z)^{-\eta},
\label{vz_profile}
\end{equation}

\noindent
where the normalization constant is defined $\mathscr{A}=\beta_0 (Z_0/Z_s)^{\eta}$,
 and $\beta_0$ is the terminal velocity at the altitude $Z_0$.

\noindent
The continuity equation for the system here considered gives the electron number density
\begin{equation}
\ne=\frac{\dot M}{\pi m_p |\beta(Z)|c R_{0}^2}, 
\label{ne}
\end{equation}

\noindent
where $\dot m \equiv \dot M/\dot M_{\rm E}$ is the mass accretion rate in Eddington units and $R_{0}$ is the radius of the accretion
column.

\noindent
We then define the adimensional quantities $z$ and $r_0$ through the change of variables $Z \rightarrow R_{\rm S\odot}mz$ 
and $R_0 \rightarrow R_{\rm S\odot} m r_0$, where $m \equiv M/M_{\odot}$,  while  $M_{\rm S\odot}$ and  $R_{\rm S\odot}$ are the Sun 
mass and Schwarzschild radius, respectively.
The effective vertical optical depth of the accretion column  is then given by
\begin{equation}
 \tau(z)=\int_{z_0}^{z} \ne \sigma_\parallel dZ'= C\frac{\dot m}{\mathscr{A}{r_0}^2}\frac{\left(z^{\eta+1}-z_{\rm 0}^{\eta+1}\right)}{\eta+1}, 
\label{tau_z}
\end{equation}

\noindent
where $C=2.2 \times 10^{-3}$, and $z_0$ is the vertical coordinate at the NS surface.

\noindent
Inverting relation (\ref{tau_z}), we also define the velocity profile of the accreting matter as a function
of the optical depth $\tau$ instead of the space variable $z$
\begin{equation}
\beta(\tau)=-\mathscr{A}\left\{z^{\eta+1}_0+ \frac{ \mathscr{A} r^2_0 (1+\eta)\tau}{C \dot m}\right\}^{-\frac{\eta}{\eta+1}}.
\label{beta_tau}
\end{equation}

\noindent
As a second possibility, following BW07, we considered the velocity profile

\begin{equation}
 \beta(\tau)=-\Psi \tau,
\label{beta_proptau}
\end{equation}

\noindent
where $\Psi=0.67\xi/z_0$ (see Eq. [32] in BW07).

\noindent
Given that in our model the optical depth $\tau$ represents one of the free parameters, once it is provided
in input together with adimensional accretion column radius $r_0$, the accretion rate $\dot{m}$ must be first
computed either from equation (\ref{tau_z}), if $\beta(\tau)$ is defined as in equation (\ref{beta_tau}),
or from equation (28) in BW07 if $\beta(\tau)$ belongs to equation (\ref{beta_proptau}).
This step is necessary to determine the $\xi$ parameter (Eq. [\ref{csi_par}]), and requires 
fixing the maximum altitude of the accretion column $z_{\rm max}$. We assumed $z_{\rm max}=2 z_0$,
and all emerging spectra (see next section) were computed with this choice.

\section{Results\label{results}}

In this section we report some examples of the theoretical spectra obtained by the numerical 
solution of equation (\ref{bweqn_aftersubs}) for different sets of the physical quantities that define the system.
We consider a BB seed photon spectrum at given temperature $\ktbb$ with exponential spatial distribution 
across the vertical direction, according to

\begin{equation}
 S(x,\tau) = C_{\rm n} e^{-\tau}  \frac{\kte^3 x^3}{e^{\kte/\ktbb~x}-1},
\end{equation}
\noindent
with the normalization constant defined as $C_{\rm n}=R^2_{\rm km}/D^2_{\rm 10}$, 
where $R_{\rm km}$ and $D_{\rm 10}$ are the BB emitting area in kilometers and
the source distance in units of 10 kpc, respectively.
The spectra were computed using the velocity profiles defined in equations
(\ref{beta_tau}) and (\ref{beta_proptau}), respectively.
The common parameters for both cases are consequently the BB temperature $\ktbb$, 
the electron temperature $\kte$, the optical depth $\tau$, the albedo at the inner
surface $A$ and the radius of the accreting column $r_0$.
On the other hand, for $\beta(\tau)$ belonging to equation (\ref{beta_tau}),
additional parameters are the index $\eta$ and the terminal velocity at the star surface $\beta_0$.
We first present the results for this second physical case.

\begin{figure*}
\includegraphics[scale=0.7]{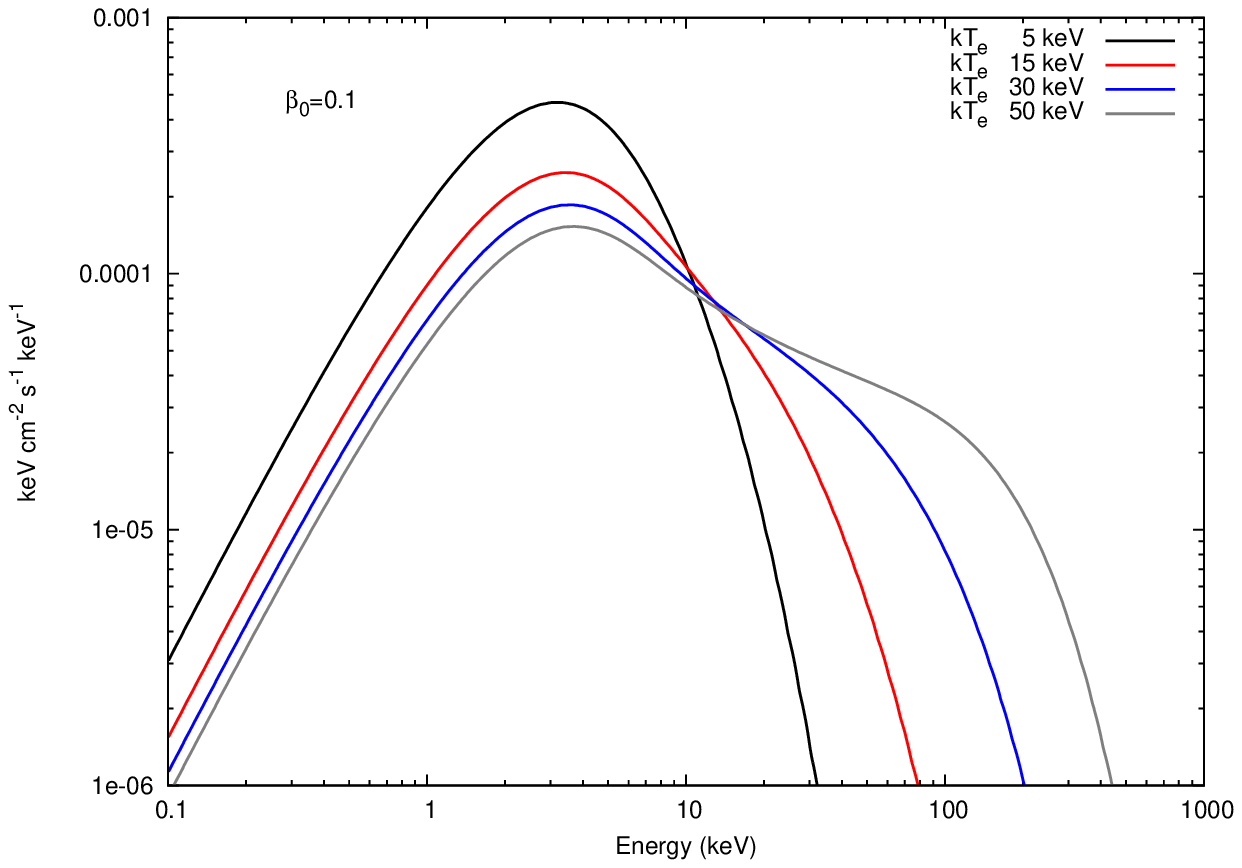}
\includegraphics[scale=0.7]{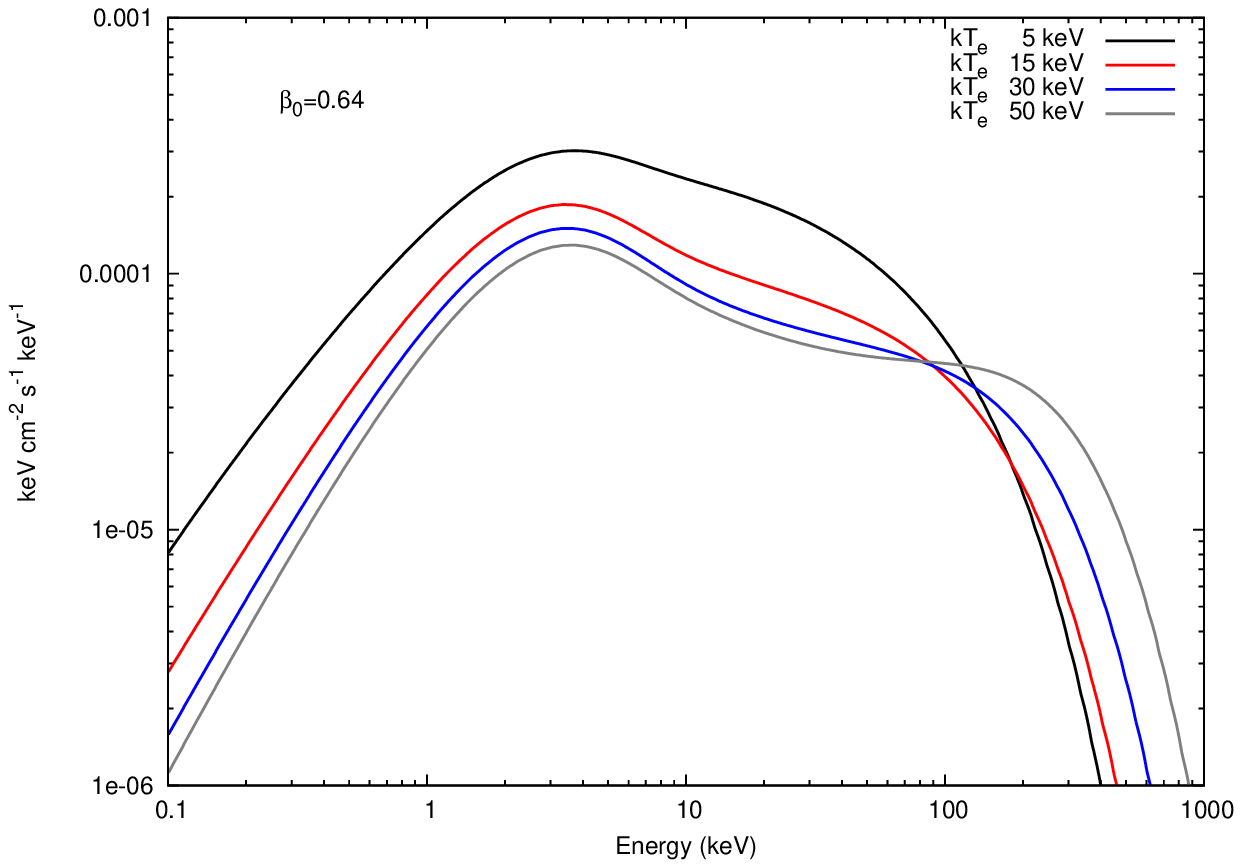}
\caption{\footnotesize{Emerging spectra obtained from the solution of equation (\ref{bweqn_aftersubs}) for different values 
of the electron temperature $\kte$ and  velocity profile defined in equation (\ref{beta_tau}). In both cases the fixed parameters are $\ktbb=1$ keV, $\tau=0.2$, $\eta=0.5$, $\r0=0.25$, $A=1$. 
\it{Left panel:} $\beta_{\rm 0}=0.1$. \it{Right panel:} $\beta_{\rm 0}=0.64$.}}
\label{fig_bw07_varkte}
\end{figure*}

In Fig. \ref{fig_bw07_varkte} we show the emerging spectra for different values of the electron temperature
$\kte$ and two terminal velocities $\beta_0=0.1$ and $\beta_0=0.64$. 
As expected, both times higher values of $\kte$ produce flatter spectra and push the cut-off energy $\Ec$ 
to higher  energies; on the other hand, the bulk contribution as a second channel of Comptonization depends 
on the value of $\kte$.
The two extreme temperature values reported here, $\kte=5$ keV and $\kte=50$ keV, are particularly instructive: 
for low electron temperatures the spectrum changes from BB-like when $\beta_0=0.1$ to a cut-off power law 
with $\Ec \ga 30$ keV  when $\beta_0=0.64$, while the spectral change is much less enhanced 
for a hot plasma. These can be considered as typical examples of
bulk-dominated and thermal-dominated Comptonization spectra, respectively.

\begin{figure*}
\includegraphics[scale=0.7]{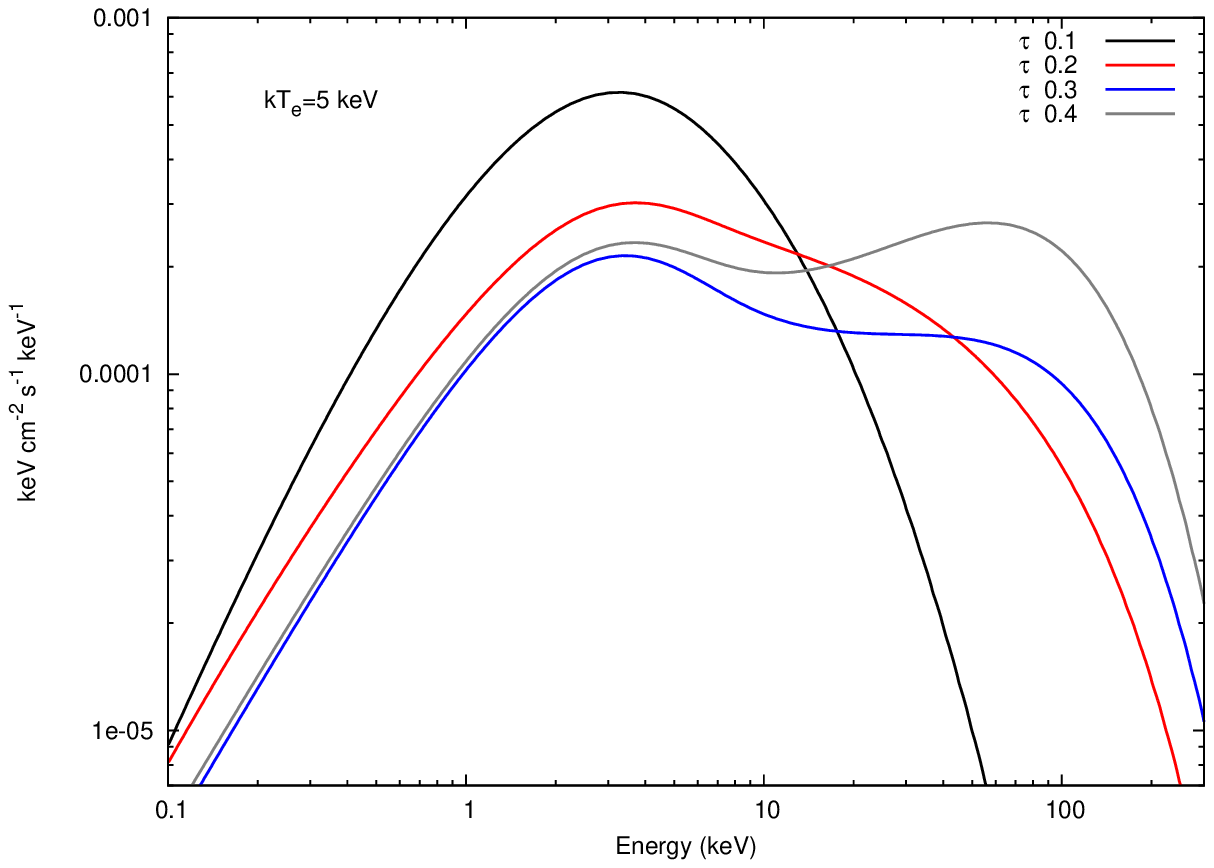}
\includegraphics[scale=0.7]{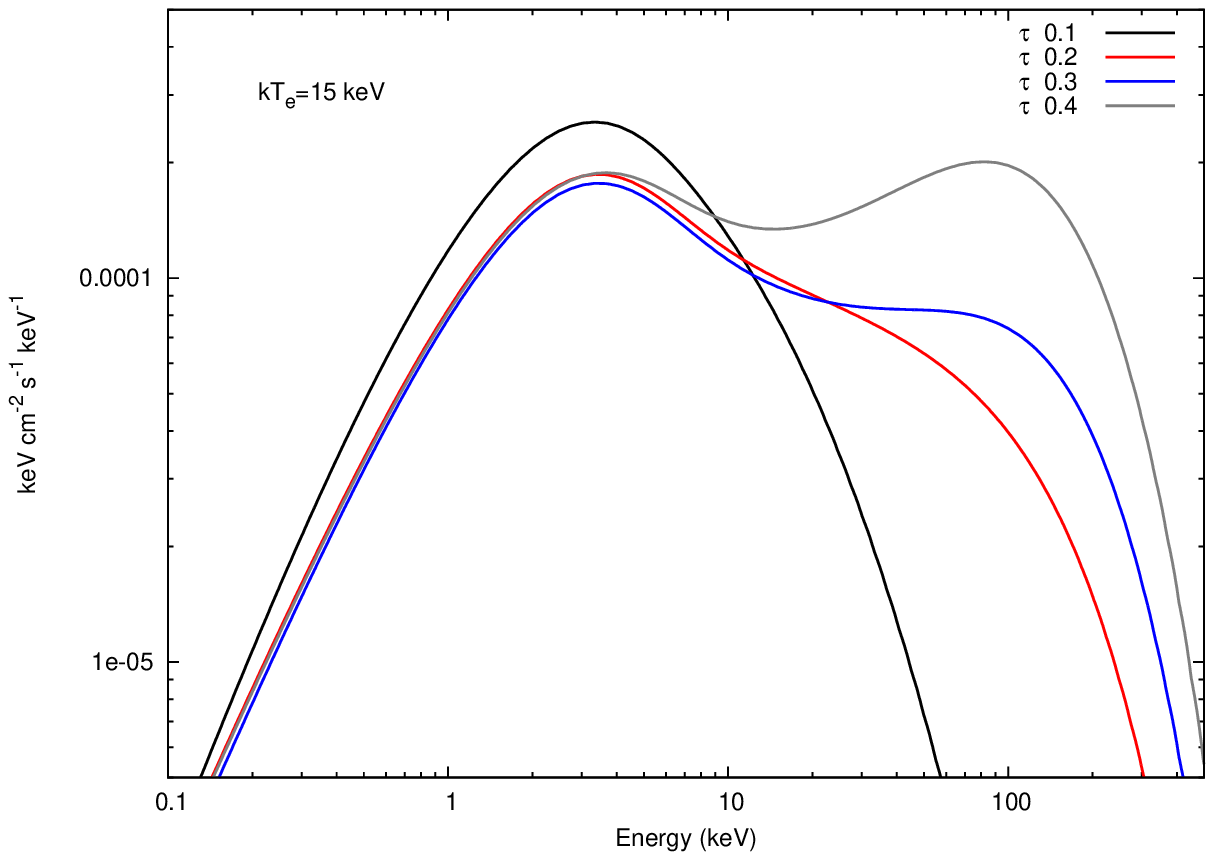}
\caption{\footnotesize{Same as Fig. \ref{fig_bw07_varkte} but for different values 
of the optical depth $\tau$. Fixed parameters are $\ktbb=1$ keV, $\eta=0.5$, $\beta_{\rm 0}=0.64$, $r_{\rm 0}=0.25$, $A=1$.
{\it Left panel:} $\kte=5$ keV. {\it Right panel:} $\kte=15$ keV.}}
\label{fig_bw07_vartau}
\end{figure*}

Together with the electron temperature, the optical depth $\tau$ is an important parameter
that plays a key role in  determining the spectral slope and cut-off energy,
as clearly shown in Fig. \ref{fig_bw07_vartau}.
We note that in Fig. \ref{fig_bw07_varkte} and Fig. \ref{fig_bw07_vartau} the index of the velocity
profile was chosen to be $\eta=0.5$, typical of accretion onto a compact object where gravity and radiation pressure 
are the only force terms that determine the dynamical configuration. 
Here, the terminal value of the matter velocity $\beta_0$ depends on the ratio of the radiative
and gravitational forces, provided the condition $|{\bf F}_{\rm r}| / |{\bf F}_{\rm g}| \la 1$ is
satisfied. This relatively simple approach is valid for low values of optical depth $\tau$,
while when $\tau > 1$  radiative transfer becomes important and the problem  requires in principle a 
more accurate radiation-hydrodynamics treatment.

\begin{figure*}
\includegraphics[scale=0.7]{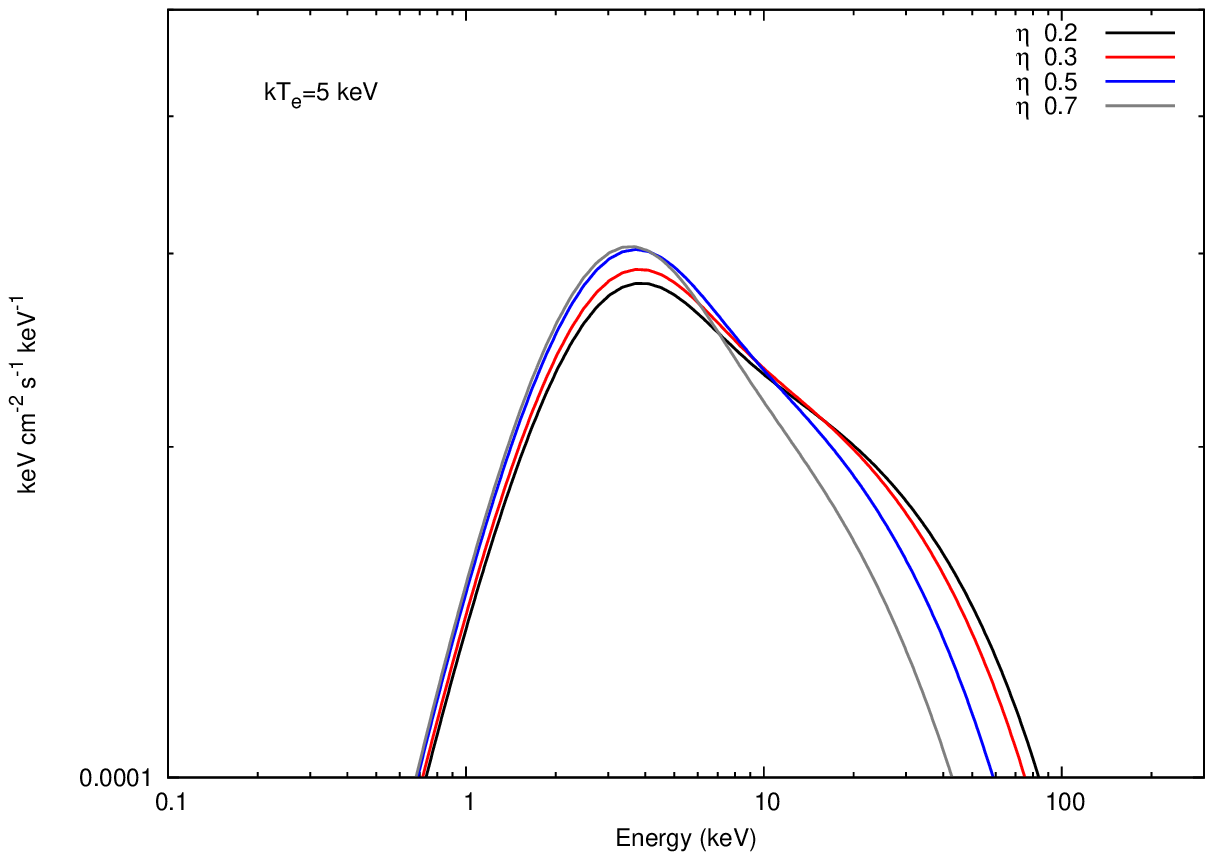}
\includegraphics[scale=0.7]{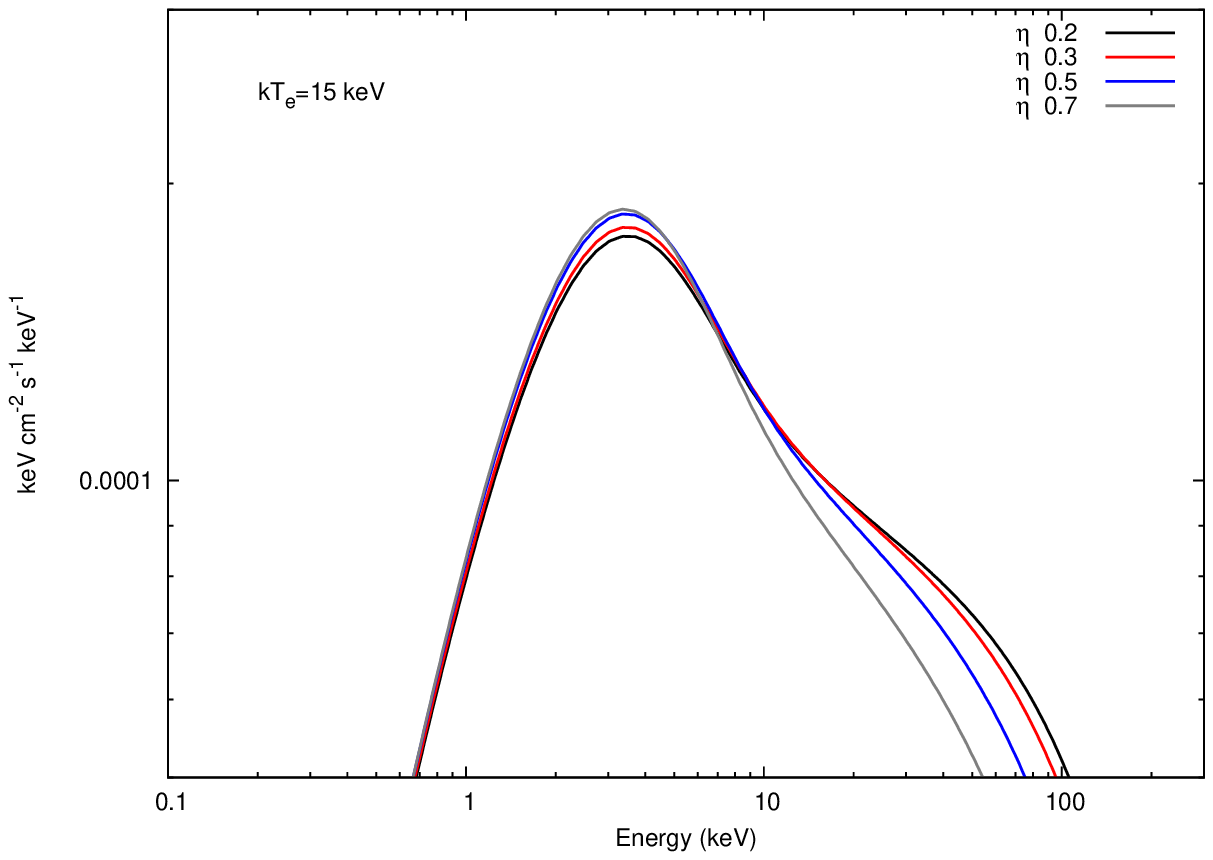}
\caption{\footnotesize{Same as Fig. \ref{fig_bw07_varkte} but for different values 
of the index of the velocity profile. Fixed parameters are $\ktbb=1$ keV, $\tau=0.2$, $\beta_{\rm 0}=0.64$,  $\r0=0.25$, $A=1$.
{\it Left panel:} $\kte=5$ keV. {\it Right panel:} $\kte=15$ keV.}}
\label{fig_bw07_varalpha}
\end{figure*}

It is outside the scope of this paper to compute the exact velocity profile for accreting matter under
the presence of a strong radiation field in a high optical depth environment. We merely introduced a simple parametrization
for modifying the velocity field by changing the index $\eta$, with
the results shown in Fig. \ref{fig_bw07_varalpha},  for two different values of the electron temperature $\kte$.
 As Fig.~\ref{fig_bw07_varalpha} shows, the lower the value of $\eta$, the harder the spectrum:  this behavior
can be explained in  a quantitative and a qualitative way. Indeed, as $\eta$ increases, the velocity profile
$\beta(z)$ becomes sharper, and for a fixed terminal velocity $\beta_0$, electron temperature $\kte$, and optical depth $\tau$, 
while photons diffuse through the bounded medium, on average the energy of the electrons (caused by their Maxwellian
plus bulk motion) will be lower,  
and consequently the net energy gain of the photons
due to inverse Compton will be less.
From the mathematical point of view, it is worth mentioning that \citet[][hereafter MK92]{mk92}  reported the analytical
solution of the RTE in the  Fokker-Planck approximation with the variable separation method for spherical accretion without magnetic field in the limit  $T_{\rm e}=0$. 
Assuming a general velocity profile $\beta_{\rm r} \propto r^{-\eta}$, the authors showed that the
spectral index of the $k^{\rm th}$-Comptonization order emerging spectrum yields $\alpha_{\rm k}=3+3\lambda_{\rm k}/(2-\eta)$ 
(see Eq. [\ref{series}]), where $\lambda_{\rm k}$ is the $k^{\rm th}$-eigenvalue of the space operator. Using
equation (B12) of MK92, it follows  immediately that as $\eta$ increases, the spectral index $\alpha_{\rm k}$ increases as well.
This mathematical result in terms of spectral formation can be considered as general in the framework of the FP 
treatment, and is accordingly qualitatively meaningful for our research.

We also emphasize that analytical solutions for $\eta \neq 0.5$ have been possible for MK92 only because of the
condition $T_{\rm e}=0$, which drops the thermal Comptonization operator in the RTE, 
while when $T_{\rm e} > 0$ this is possible only for $\eta=0.5$ (TMK97, F08).

\begin{figure*}
\includegraphics[scale=0.7]{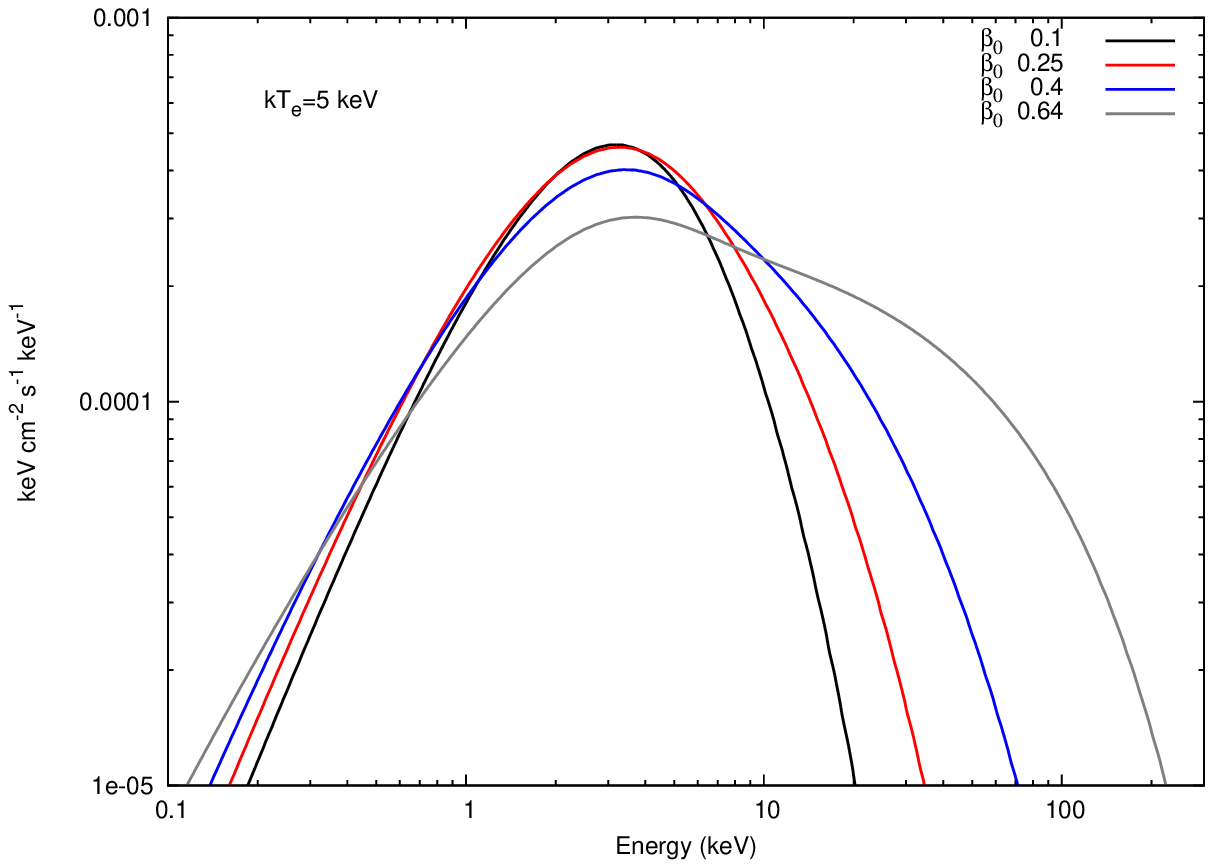}
\includegraphics[scale=0.7]{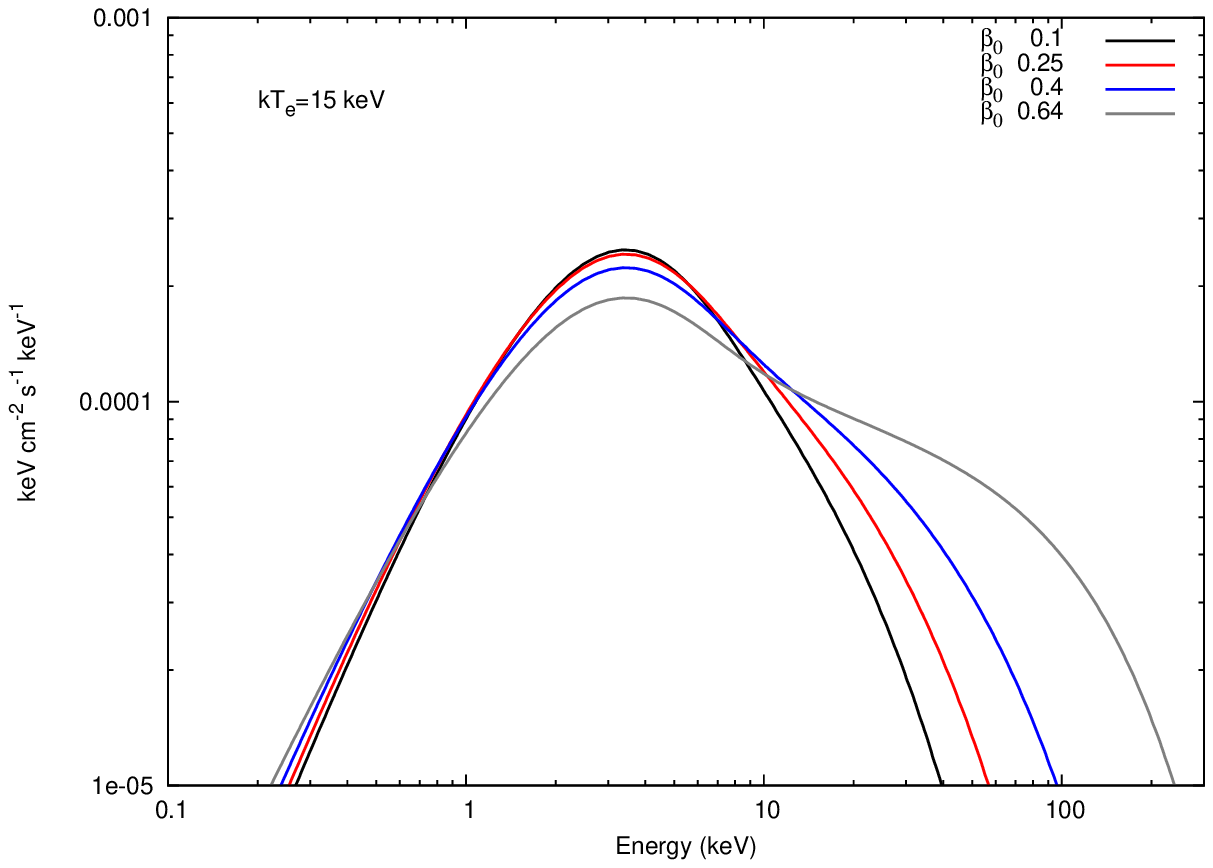}
\caption{\footnotesize{Same as Fig. \ref{fig_bw07_varkte} but for different values 
of the inner velocity $\beta_{\rm 0}$. Fixed parameters are $\ktbb=1$ keV, $\tau=0.2$, $\eta=0.5$,  $\r0=0.25$, $A=1$.
{\it Left panel:} $\kte=5$ keV. {\it Right panel:} $\kte=15$ keV.}}
\label{fig_bw07_varvel}
\end{figure*}

In Fig.~\ref{fig_bw07_varvel} we show results for different terminal bulk velocities $\beta_0$ for two electron temperature values. The figure can be considered an extension and completion of 
Fig.~\ref{fig_bw07_varkte} because more values of $\beta_0$ are shown to better appreciate the induced changes in the emerging spectra.

\begin{figure*}
\includegraphics[scale=0.7]{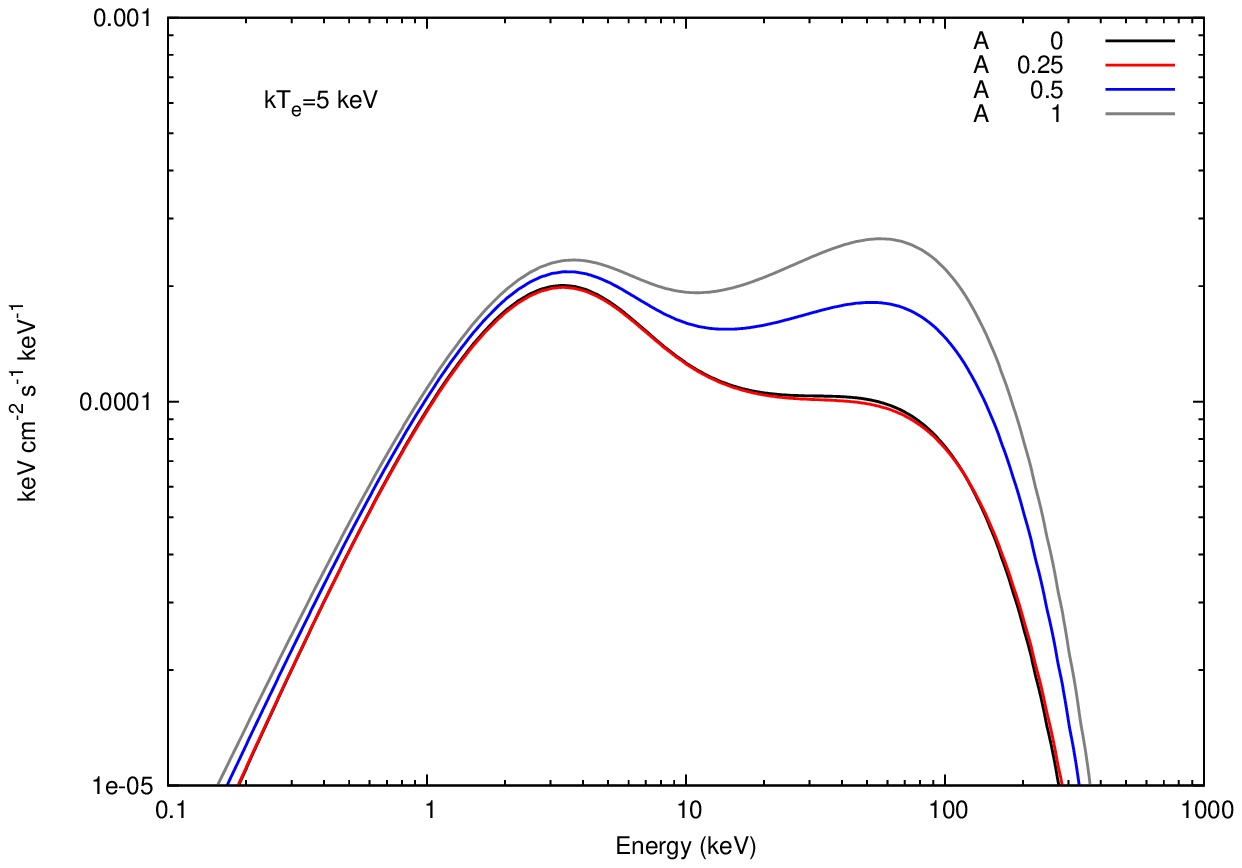}
\includegraphics[scale=0.7]{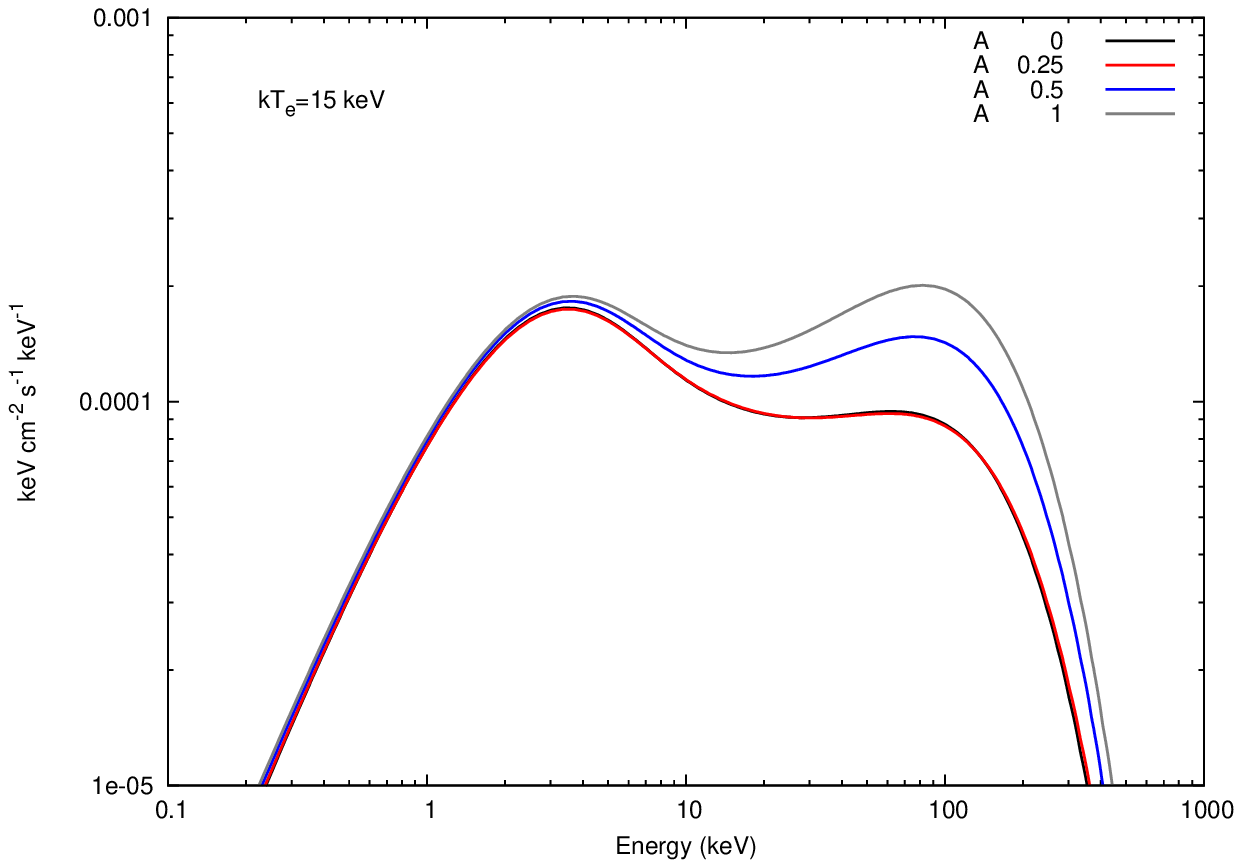}
\caption{\footnotesize{Same as Fig. \ref{fig_bw07_varkte} but for different values 
of the albedo $A$, with the velocity profile of equation (\ref{beta_tau}). Fixed parameters are $\ktbb=1$ keV, $\tau=0.4$, $\eta=0.5$, $\beta_{\rm 0}=0.64$, $\r0=0.25$.
{\it Left panel:} $\kte=5$ keV, {\it Right panel:} $\kte=15$ keV.}}
\label{fig_bw07_varalbedo}
\end{figure*}

The spectral modifications as a result of different values of the albedo $A$ at the inner surface
are instead shown in Fig. \ref{fig_bw07_varalbedo}, where we explored  full absorption ($A=0$) and full reflection ($A=1$), together with other intermediate values.
In the framework of a physical link to astrophysical objects it would be natural to
associate a BH to the condition $A=0$ and  a NS to the condition $A=1$, respectively \citep{tf04,ft11},
even though this latter assumption may be considered an oversimplification of the problem.
A most realistic approach would consist indeed in an energy-dependent treatment of the albedo, a problem
that could be faced only with Montecarlo simulations, with the additional complications arising
from a detailed treatment of the star photosphere (surface) properties.

For our unavoidably simplified assumptions, the net effect of increasing values of
$A$ is a progressive flattening of the emerging spectra. This is physically explained because when $A> 0$, a 
fraction of photons (which becomes 100\% when A=1) suffers on average more scattering
with respect to  $A=0$. Qualitatively, the spectral modification leads in the
same direction as an enhanced optical depth of the system.

\begin{figure*}
\includegraphics[scale=0.7]{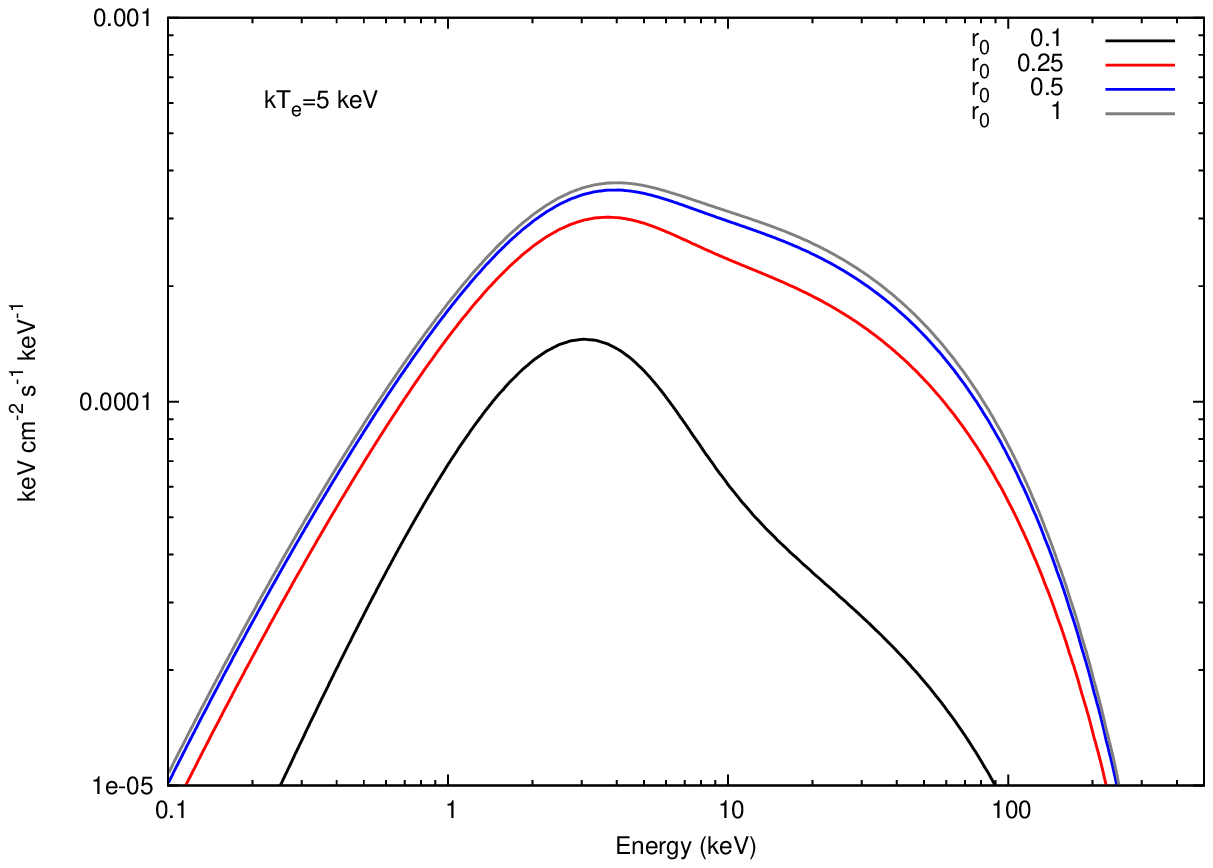}
\includegraphics[scale=0.7]{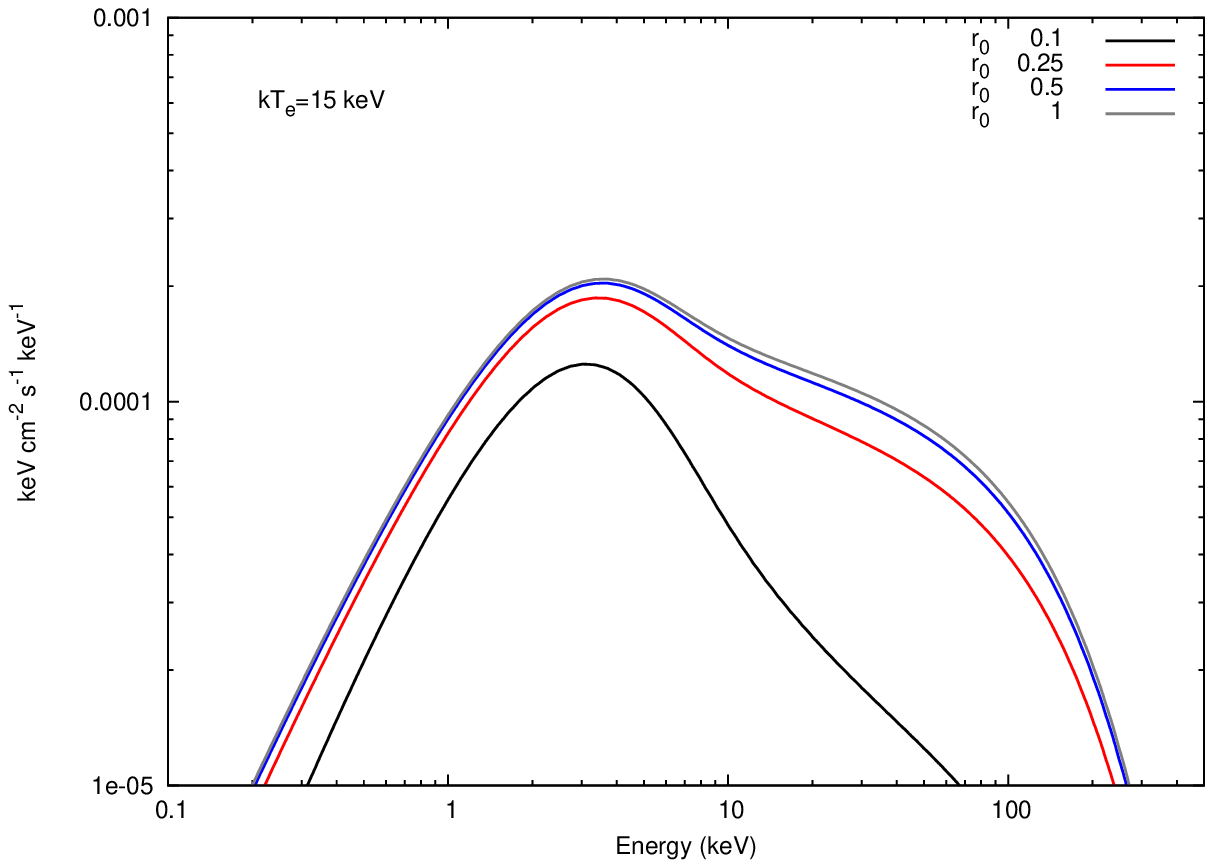}
\caption{\footnotesize{Same as Fig. \ref{fig_bw07_varkte} but for different values 
of the accretion column radius $\r0$, with the velocity profile of equation (\ref{beta_tau}). Fixed parameters are $\ktbb=1$ keV, $\tau=0.2$, $\eta=0.5$, $\beta_{\rm 0}=0.64$, $A=1$.
{\it Left panel:} $\kte=5$ keV. {\it Right panel:} $\kte=15$ keV.}}
\label{fig_bw07_varr0}
\end{figure*}

The last  parameter that strongly influences the spectral formation is the radius
of the accretion column $r_0$, whose effects are shown in Fig. \ref{fig_bw07_varr0}.
Indeed, following the BW07 prescription,  the mean escape time for photons 
using the diffusion approximation  $t_{\rm esc} \propto r^2_0$ (see Eq. [\ref{tesc}]).
On the other hand, both the bulk and thermal Comptonization  parameters ($y_{\rm bulk}$ and $y_{\rm th}$, respectively)
are related to the mean number of scatterings that photons experience in the medium via
\begin{eqnarray}
y_{\rm bulk} &\approx& N^{\rm bulk}_{\rm av} \zeta_{\rm bulk},\\\nonumber
y_{\rm th} &\approx& N^{\rm th}_{\rm av} \zeta_{\rm th},
\label{y_par} 
\end{eqnarray}

\noindent
where $N^{\rm bulk}_{\rm av}$, $\zeta_{\rm bulk}$, $N^{\rm th}_{\rm av}$ and $\zeta_{\rm th}$
are the averaged number of scatterings and the fraction energy gain per scattering for bulk and thermal
Comptonization, respectively. Both $N^{\rm bulk}_{\rm av}$ and $N^{\rm th}_{\rm av}$ are 
of course also proportional to $t_{\rm esc}$ (see Eqs.[94]-[97] in BW07).

Evidently therefore, for fixed velocity profile parameters $\mathscr{A}$ and $\eta$ (see Eq. [\ref{vz_profile}]),
once the optical depth $\tau$ is defined (see Eq. [\ref{tau_z}]), to keep its value constant for increasing  $r_0$
(as reported in Fig. \ref{fig_bw07_varr0}), 
the accretion rate $\dot{m}$ must also increase in a way to keep the ratio $\dot{m}/r^2_0$ constant.
Combining equation (\ref{tesc}) and (\ref{ne}) yields $t_{\rm esc} \propto \dot{m}$, which in turn leads to an enhancement of the Comptonization parameters $y_{\rm bulk}$ and $y_{\rm th}$ in equation (\ref{y_par}) with a hardening of the spectral shape.

\begin{figure*}
\includegraphics[scale=0.7]{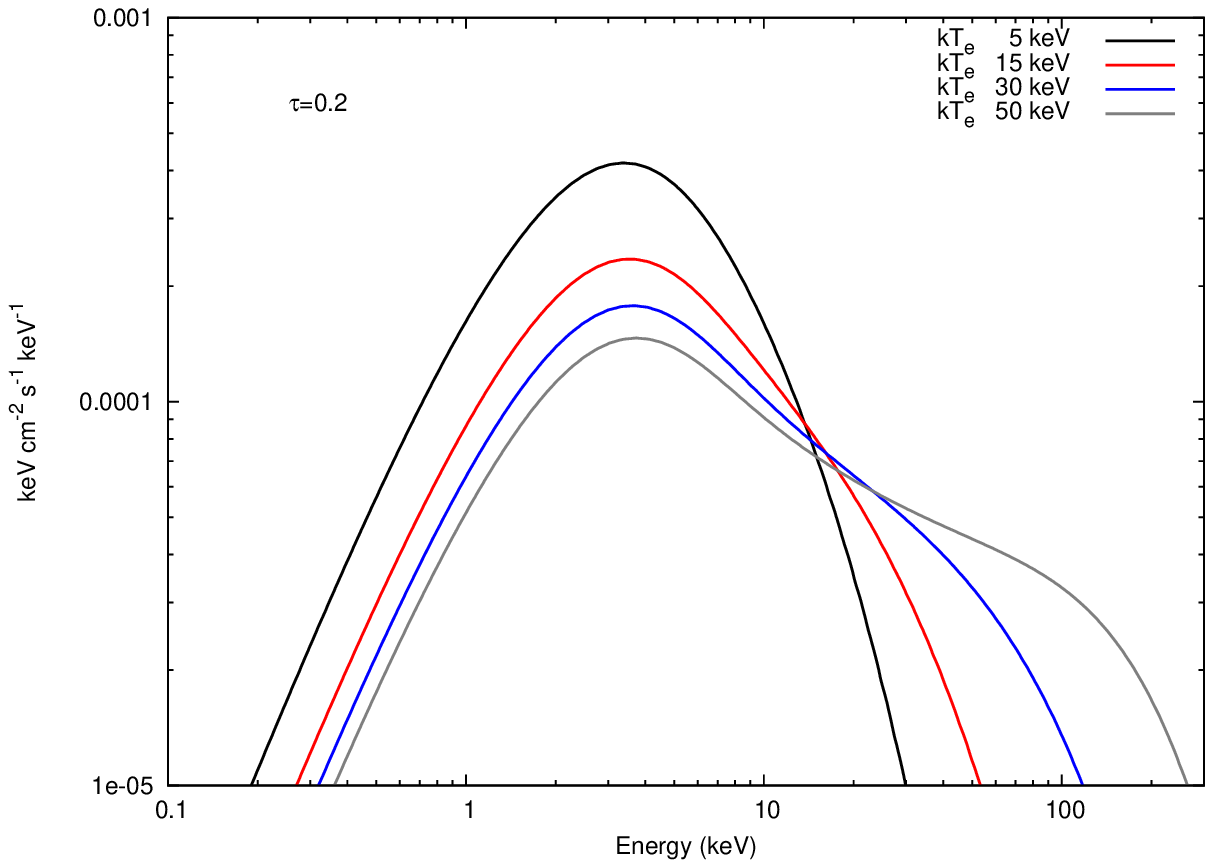}
\includegraphics[scale=0.7]{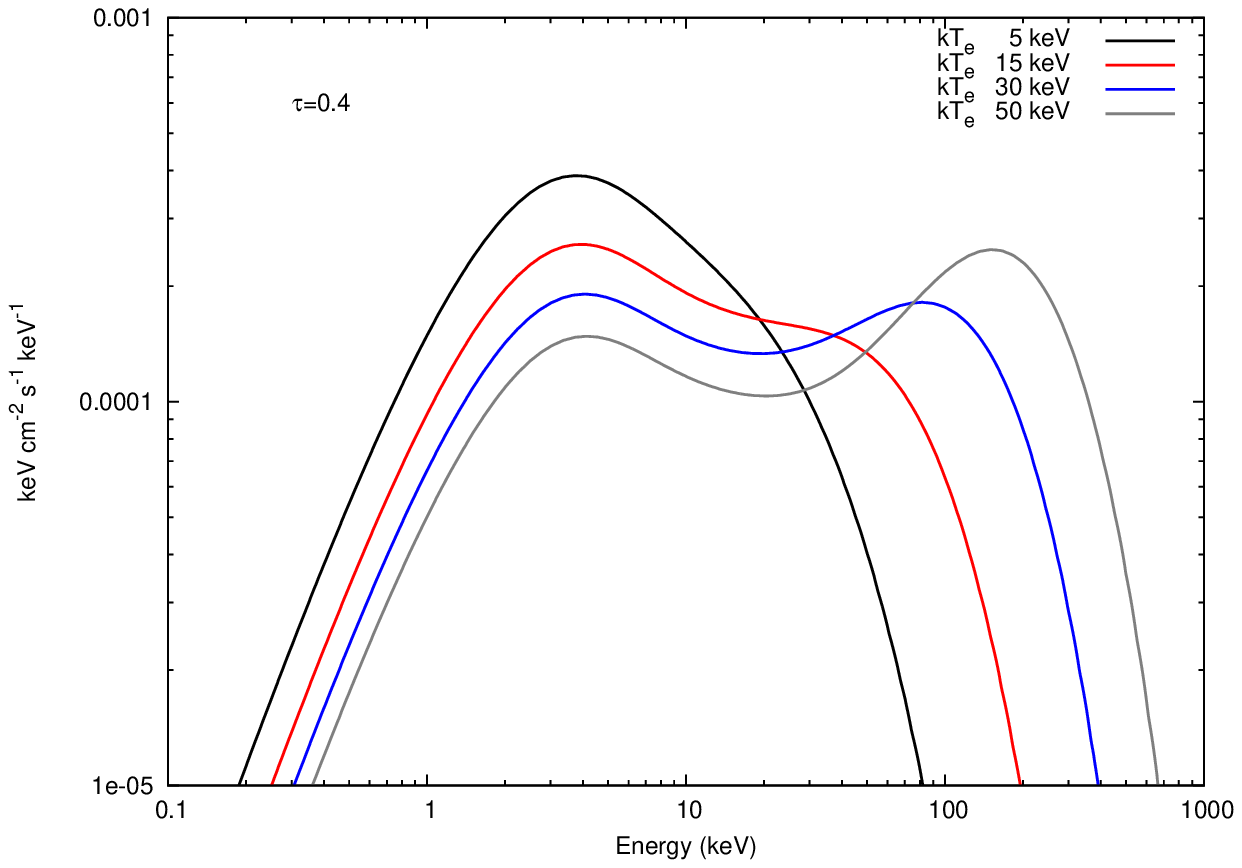}
\caption{\footnotesize{Emerging spectra obtained from the solution of equation (\ref{bweqn_aftersubs}) for different values 
of the electron temperature $\kte$, with the velocity profile of equation (\ref{beta_proptau}). In both cases the fixed parameters are $\ktbb=1$ keV, $\r0=0.25$, $A=1$. \it{Left panel:} $\tau=0.2$. \it{Right panel:} $\tau=0.4$.}}
\label{fig_bw07_varkte_prof1}
\end{figure*}

\begin{figure*}
\includegraphics[scale=0.7]{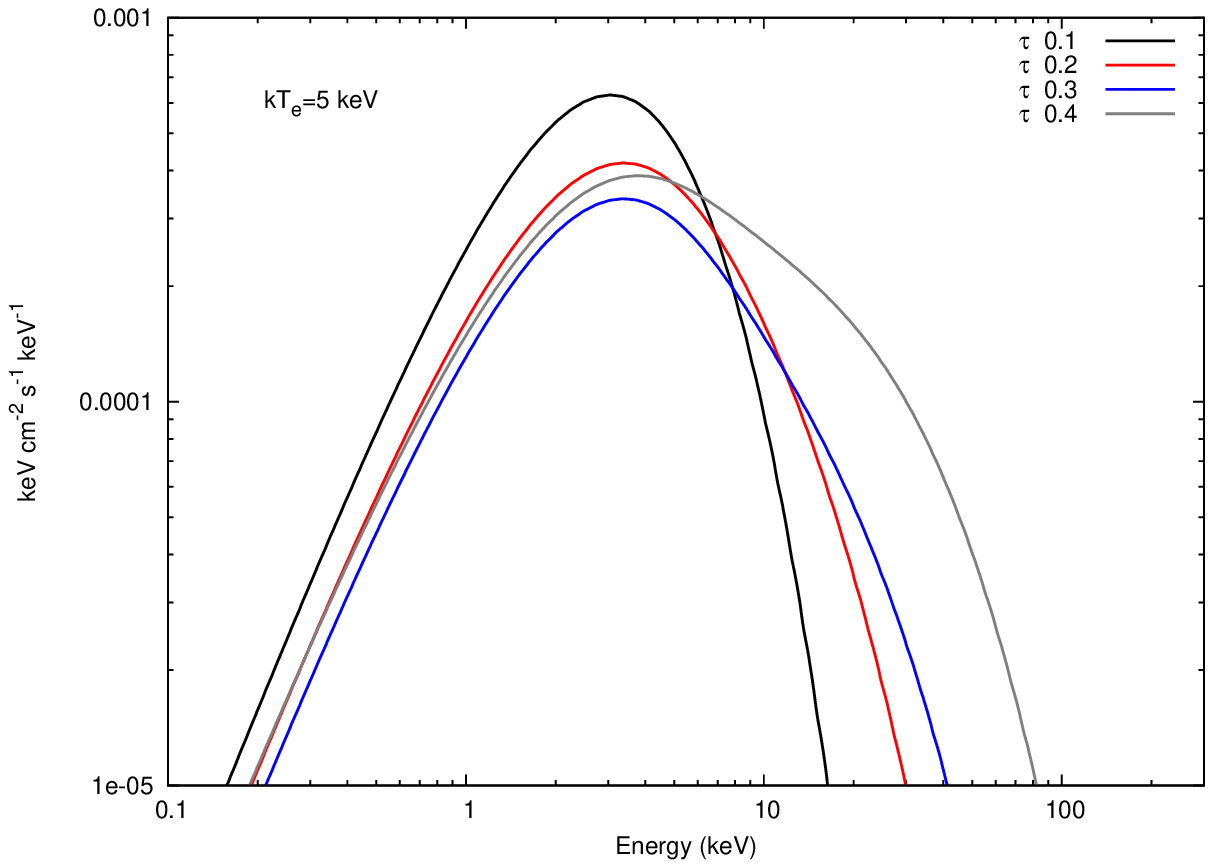}
\includegraphics[scale=0.7]{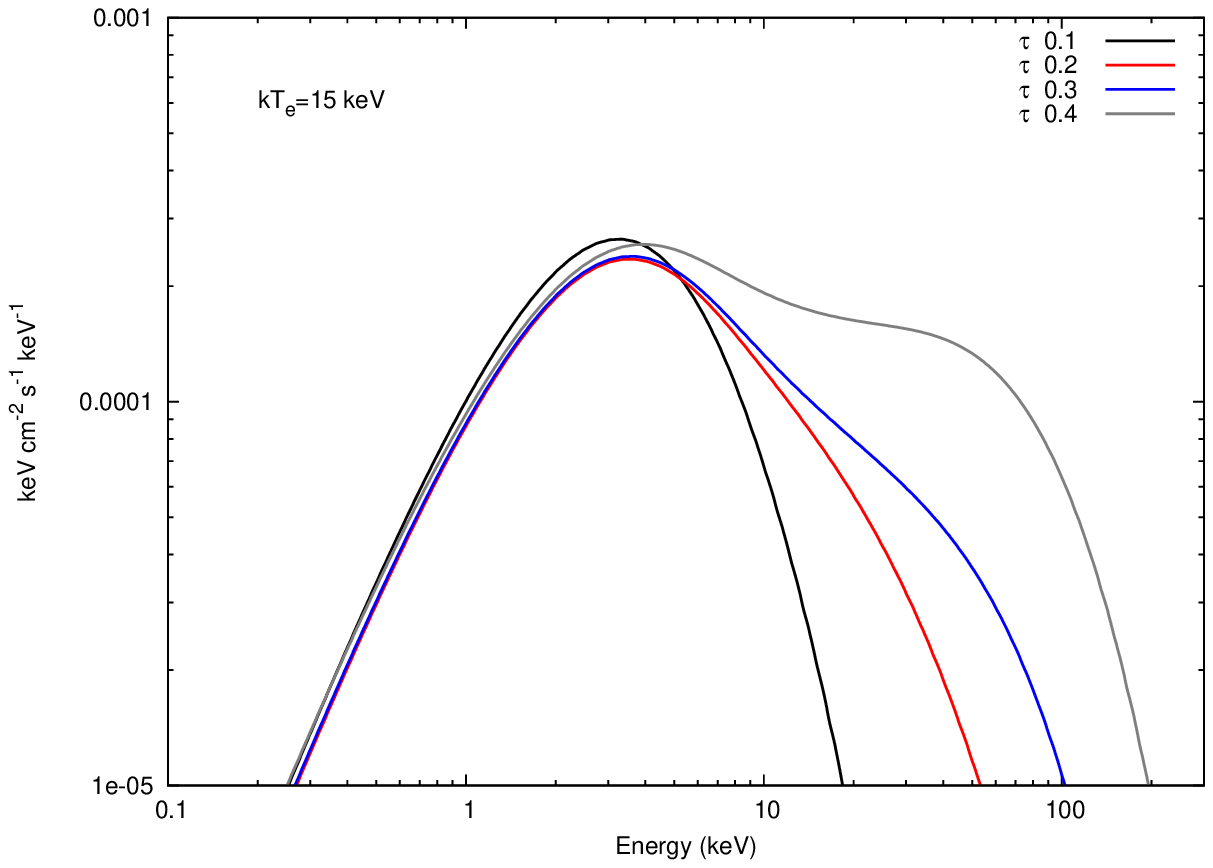}
\caption{\footnotesize{Same as Fig. \ref{fig_bw07_varkte} but for different values 
of the optical depth $\tau$, with the velocity profile of equation (\ref{beta_proptau}). 
Fixed parameters are $\ktbb=1$ keV, $r_{\rm 0}=0.25$, and $A=1$.
{\it Left panel:} $\kte=5$ keV. {\it Right panel:} $\kte=15$ keV.}}
\label{fig_bw07_vartau_prof1}
\end{figure*}

\begin{figure*}
\includegraphics[scale=0.7]{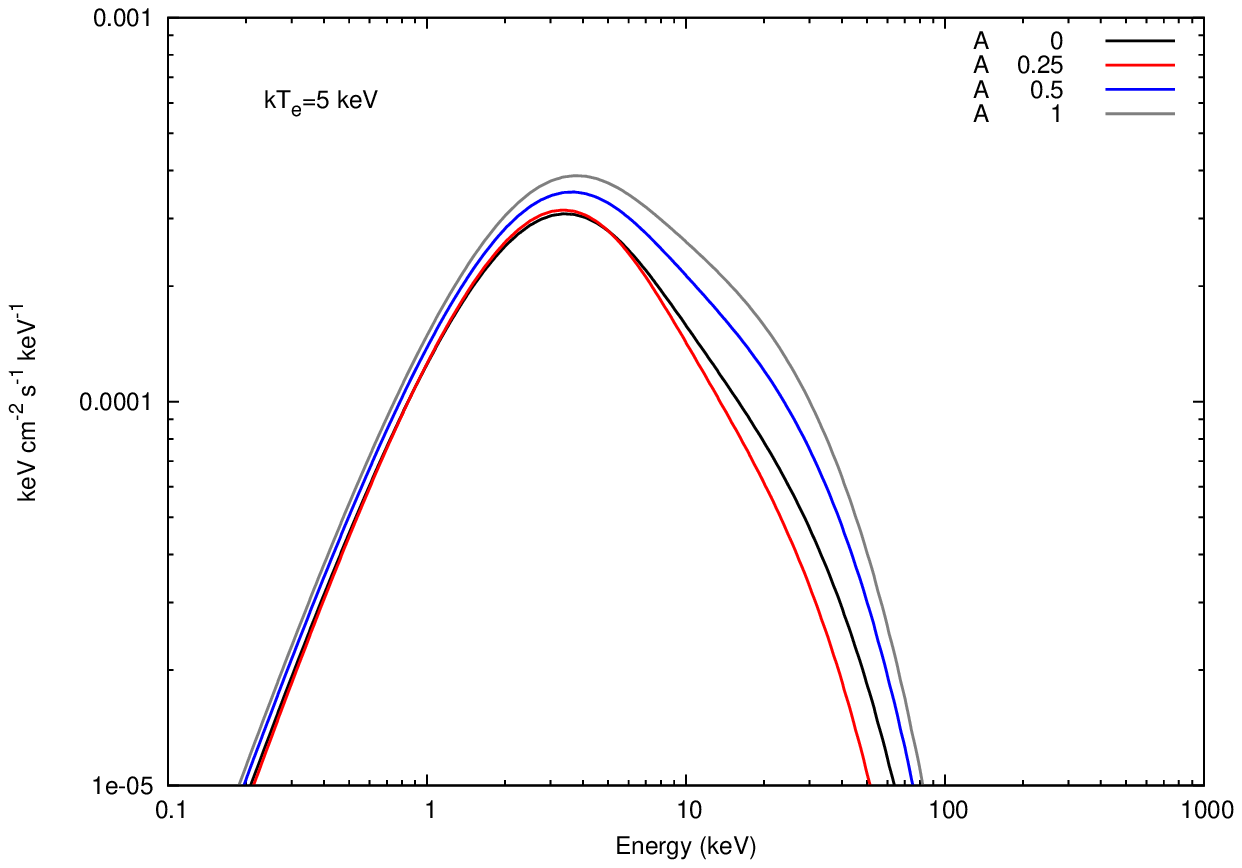}
\includegraphics[scale=0.7]{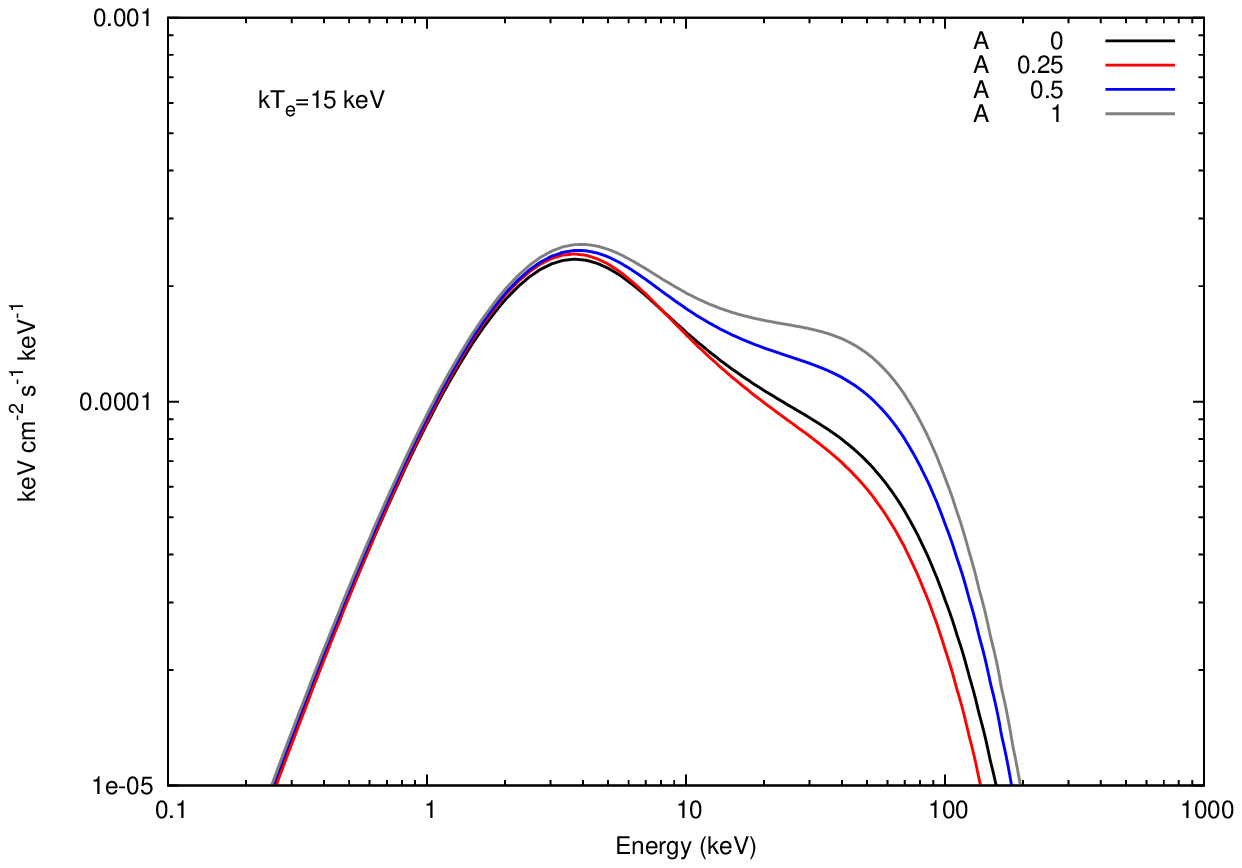}
\caption{\footnotesize{Same as Fig. \ref{fig_bw07_varkte} but for different values 
of the albedo $A$, with the velocity profile of equation (\ref{beta_proptau}). Fixed parameters are $\ktbb=1$ keV, 
$\tau=0.4$, and $\r0=0.25$.
{\it Left panel:} $\kte=5$ keV. {\it Right panel:} $\kte=15$ keV.}}
\label{fig_bw07_varalbedo_prof1}
\end{figure*}

Considering now the velocity profile defined in equation (\ref{beta_proptau}), we see that  the results 
are qualitatively the same as in equation (\ref{beta_tau})  as far as the spectral
modifications induced by variations of $\kte$ are concerned (Fig. \ref{fig_bw07_varkte_prof1}), 
$\tau$ (Fig. \ref{fig_bw07_vartau_prof1}) and $A$ (Fig. \ref{fig_bw07_varalbedo_prof1}), respectively.
But there are opposite effects that are induced in the emerging spectra by different
values of the accretion column radius $r_0$ for the velocity profile here considered.
Indeed, using equation (\ref{csi_par}) and  the definition of $\tau$ in equation (28) of BW07,
which allows us to express the accreting matter velocity in terms of the $z$-coordinate, in spite
of the optical depth $\tau$, it is straightforward to see that $\beta(z) \propto r^{-1/2}_0$.
In particular, if $z_0=2.42$ and $z_{\rm max}=2z_0$ we have  $\beta_{\rm max}$=0.60 for $r_0=0.1$,
$\beta_{\rm max}$=0.38 for $r_0=0.25$, $\beta_{\rm max}$=0.27 for $r_0=0.5$ and  $\beta_{\rm max}$=0.2
for $r_0=1$, respectively. 
Note that because equation (\ref{beta_proptau}) describes matter that stagnates
at the star surface,  here $\beta_{\rm max}$ represents the velocity at the 
accretion column altitude $z_{\rm max}$.
In other words, while using equation (\ref{beta_tau}), the choice of $r_0$ does not modify
the velocity field of the accreting matter, which is only determined by the choice of $\beta_0$
and $\eta$, for (\ref{beta_proptau}) as $r_0$ increases the bulk contribution
to the spectral formation becomes less important, and this drop is not compensated for
 by the increase of the photon mean escape time $t_{\rm esc}$, which, as explained above,
would instead contribute to spectral hardening.

\section{Implementation in the \xspec\ package}
\label{4xspec}
 \begin{table}  
 \begin{center}      
\tabcolsep 4pt   
 \caption{Parameter description of the \xspec\ model \modname. \label{xspecpars}}  
 \begin{tabular}{lll} 
 \hline 
 \hline 
 \noalign{\smallskip} 
Parameter & Units       &  Description        \\
 \noalign{\smallskip} 
 \hline 
 \noalign{\smallskip} 
$kT_{\rm bb}$        &(keV)  &  Seed photon blackbody temperature\\
$kT_{\rm e}$         &(keV)  &  Electron temperature \\
$\tau$              &   &	  Optical depth of the accretion column \\
$\eta$              &	&	 Index of the velocity profile  (Eq. [\ref{beta_tau}])\\
$\beta_0$           &   &        Terminal velocity at the NS surface  \\
                    &   &         (Eq. [\ref{beta_tau}])   \\
$r_0$	            &   &	  Radius of the accretion column in\\
	            &   &	  units of the NS \scw\ radius\\
$A$	            &   &	  Albedo at the NS surface \\
Flag                &   &         $=1$, $\beta(\tau)$ from equation (\ref{beta_tau}) \\ 
                    &   &         $=2$, $\beta(\tau)$ from equation (\ref{beta_proptau}) \\ 
Norm	            &   &	  $R^2_{\rm km}/D^2_{10}$\\
  \noalign{\smallskip}
  \hline
  \end{tabular}
  \end{center}
  \end{table}

Our model will be publically available and distributed as a contributed model 
to the official \xspec\footnote{http://heasarc.nasa.gov/xanadu/xspec/newmodels.html} 
web page.\\ 
In Table \ref{xspecpars} we report a summary of the free parameters of the model with their physical meaning.
The code is written using C-language, and can be easily installed following
the standard procedure reported in the official \xspec\ manual and in the brief cookbook, which will be delivered together with the source code.
As a general concern for users, we point out that usually the
emerging spectrum obtained from the Comptonization of a seed photon
population with any given energy distribution S(E) can be presented
as the sum of the seed spectrum and its convolution with
the scattering Green's function $ G(E,E_0)$ of the electron plasma, 
each  with their relative weight, according to the general formalism
\begin{equation}
 F(E)= \frac{C_{\rm n}}{A+1} [S(E)+ A \times S(E) \ast G(E,E_0)],
\label{bmc}
\end{equation}

\noindent
where $C_{\rm n}$ is a normalization constant. The ratio
$A/(A+1)$ is the Comptonization fraction, and its value qualitatively
determines the contribution to the total spectrum of
the Comptonized photons. 
The value of $A$, here not to be confused with the albedo, may depend on several geometrical and physical
factors, such as the spatial seed photon distribution
inside the system configuration (see Fig. 4 in TMK97).
The lower the value of $A$, the more enhanced  the direct seed photon spectrum S(E).
Examples of \xspec\ models that use the definition in equation (\ref{bmc})
are \bmc\ (TMK97) and \comptb\ (F08).
Either model, however, does not solve the full RTE including the 
photon spatial diffusion and distribution, the latter of which is an
unknown quantity that is phenomenologically described through the
continuum parameter log(A).
On the other hand, it is not possible to change the value of log(A) arbitrarily in our present model, i.e., 
according to the observed spectra. Its value is implicitly determined
once the seed photon spatial distribution is fixed. 

We presented the results of simulated spectra assuming an exponential distribution
over $\tau$ for S(E), which was assumed to be a BB; then, the transition from the low-energy part of the spectrum
(the Rayleigh regime for $E \la 3 \ktbb$) to the high-energy (Comptonized) powerlaw 
shape is almost smooth, which  corresponds approximately to $A \gg 1$ in equation
(\ref{bmc}).
Other seed photons spatial distributions  can produce a different onset between the BB peak and 
the powerlaw-like regime. In general, for observed spectra where a direct and enhanced BB-like
component is required by the fit, our claim is to model the source
continuum with modelization of the type {\sc bb} + \modname\ by 
preferably keeping equal to each other the temperatures of the direct and Comptonized BB component.

\section{Conclusions}

We have developed a numerical code for solving  elliptic partial differential
equations based on a relaxation method with finite differences.
In particular, we reported a specific application of the algorithm to the radiative
transfer equation in the Fokker-Planck approximation, which is of particular
interest for high-energy astrophysical applications. We considered cylindrical accretion onto the polar cap of a magnetized neutron
star, using the mathematical formulation of the problem reported in 
\cite{bw07} with some modifications. Specifically, we included the second-order bulk Comptonization term in the
scattering operator and we considered different velocity profiles for the accreting matter.

The code for the computation of the emerging spectra in this configuration
has been written with the aim to implement it in the \xspec\ package
and will be delivered to the scientific community.
Because angle-averaged cross-sections caused by the magnetic field were included, the model is suitable
to be applied to the observed spectra of sources where
most of spectral formation is claimed to form
close to a NS with high magnetic field ($B \ga 10^{12}$ G),
such as accreting \mbox{X-ray} pulsars and supergiant fast \mbox{X-ray} transients.
Of course there are some unavoidable simplifications in the model, such as the assumption
of constant electron temperature of the accretion column. We note, however, that 
the isothermal condition is typical of any Comptonization model implemented
in XSPEC because it is fundamental for users  to reach a compromise between 
the accuracy of the physical treatment and the computational speed. 
More accurate theoretical investigations of the accretion processes
can be performed only with MHD simulations  performed through PC cluster-computing resources, 
 which are beyond the scope of the present work.

As in many theoretical models, the number
of available free parameters is higher than the number of
the observable ones. Therefore a correct working approach is to
keep some parameters fixed during the \mbox{X-ray} spectral fitting procedure to avoid degeneracy.

If this model is used in any publication, we kindly ask the authors to 
cite this paper in the reference list.

\begin{figure*}
\includegraphics[scale=0.7]{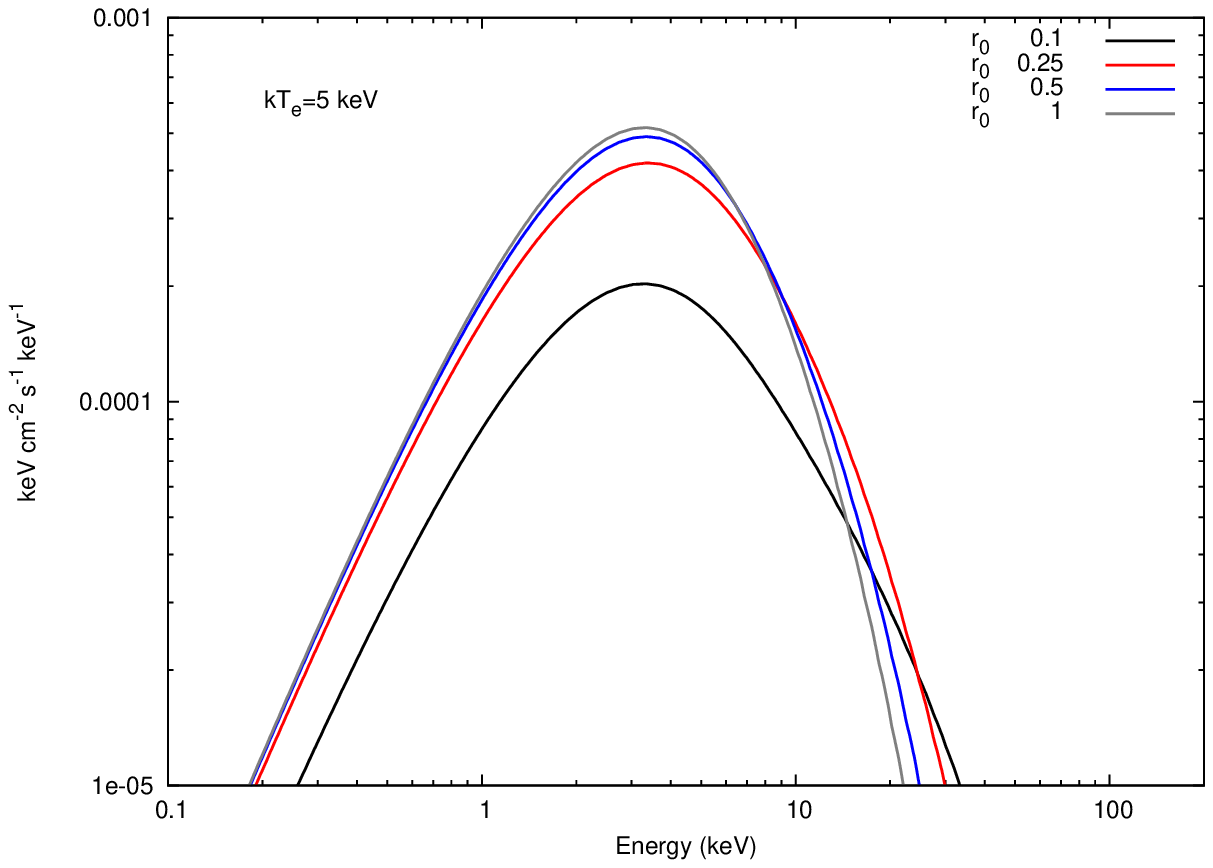}
\includegraphics[scale=0.7]{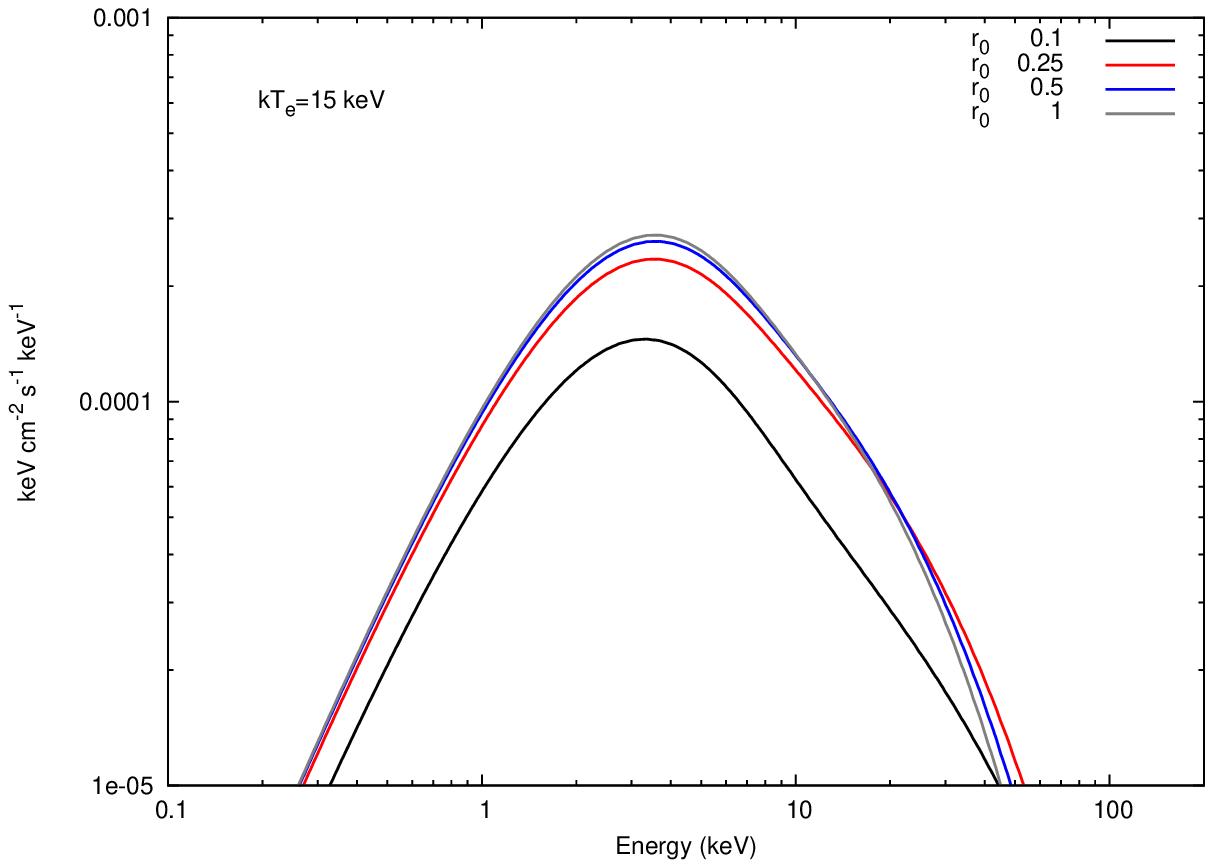}
\caption{\footnotesize{Same as Fig. \ref{fig_bw07_varkte} but for different values 
of the accretion column radius $\r0$, with the velocity profile of equation (\ref{beta_proptau}). 
Fixed parameters are $\ktbb=1$ keV, $\tau=0.2$, and $A=1$. 
{\it Left panel:} $\kte=5$ keV. {\it Right panel:} $\kte=15$ keV.}}
\label{fig_bw07_varr0_prof1}
\end{figure*}

\acknowledgements

We acknowledge financial contribution from the agreement ASI-INAF I/009/10/0.

\newpage

\bibliographystyle{aa}
\bibliography{biblio}

\begin{thebibliography}{14}
\expandafter\ifx\csname natexlab\endcsname\relax\def\natexlab#1{#1}\fi

\bibitem[{{Becker} \& {Wolff}(2007)}]{bw07}
{Becker}, P.~A. \& {Wolff}, M.~T. 2007, \apj, 654, 435 (BW07)

\bibitem[{{Blandford} \& {Payne}(1981)}]{bp81a}
{Blandford}, R.~D. \& {Payne}, D.~G. 1981, \mnras, 194, 1041 (BP81)

\bibitem[{{Farinelli} \& {Titarchuk}(2011)}]{ft11}
{Farinelli}, R. \& {Titarchuk}, L. 2011, \aap, 525, A102

\bibitem[{{Farinelli} {et~al.}(2008){Farinelli}, {Titarchuk}, {Paizis}, \&
  {Frontera}}]{f08}
{Farinelli}, R., {Titarchuk}, L., {Paizis}, A., \& {Frontera}, F. 2008, \apj,
  680, 602 (F08)

\bibitem[{{Ferrigno} {et~al.}(2009){Ferrigno}, {Becker}, {Segreto}, {Mineo}, \&
  {Santangelo}}]{ferrigno09}
{Ferrigno}, C., {Becker}, P.~A., {Segreto}, A., {Mineo}, T., \& {Santangelo},
  A. 2009, \aap, 498, 825

\bibitem[{{Lyubarskii} \& {Sunyaev}(1982)}]{ls82}
{Lyubarskii}, Y.~E. \& {Sunyaev}, R.~A. 1982, Soviet Astronomy Letter, 8, 330

\bibitem[{{Mastichiadis} \& {Kylafis}(1992)}]{mk92}
{Mastichiadis}, A. \& {Kylafis}, N.~D. 1992, \apj, 384, 136 (MK92)

\bibitem[{{Pomraning}(1973)}]{pomraning73}
{Pomraning}, G.~C. 1973, {The equations of radiation hydrodynamics}, ed.
  {Pomraning, G.~C.}

\bibitem[{{Press} {et~al.}(1992){Press}, {Teukolsky}, {Vetterling}, \&
  {Flannery}}]{NR}
{Press}, W.~H., {Teukolsky}, S.~A., {Vetterling}, W.~T., \& {Flannery}, B.~P.
  1992, {Numerical recipes in C. The art of scientific computing}, ed. {Press,
  W.~H., Teukolsky, S.~A., Vetterling, W.~T., \& Flannery, B.~P. }

\bibitem[{{Rybicki} \& {Lightman}(1979)}]{rl79}
{Rybicki}, G.~B. \& {Lightman}, A.~P. 1979, {Radiative processes in
  astrophysics}, ed. {Rybicki, G.~B.~\& Lightman, A.~P.}

\bibitem[{{Titarchuk} \& {Fiorito}(2004)}]{tf04}
{Titarchuk}, L. \& {Fiorito}, R. 2004, \apj, 612, 988

\bibitem[{{Titarchuk} \& {Lyubarskij}(1995)}]{tl95}
{Titarchuk}, L. \& {Lyubarskij}, Y. 1995, \apj, 450, 876

\bibitem[{{Titarchuk} {et~al.}(1997){Titarchuk}, {Mastichiadis}, \&
  {Kylafis}}]{tmk97}
{Titarchuk}, L., {Mastichiadis}, A., \& {Kylafis}, N.~D. 1997, \apj, 487, 834
  (TMK97)

\bibitem[{{Titarchuk} \& {Zannias}(1998)}]{tz98}
{Titarchuk}, L. \& {Zannias}, T. 1998, \apj, 493, 863

\end{thebibliography}

\end{document}